\documentclass[12pt]{article}

\usepackage[margin=1in]{geometry}
\usepackage{amssymb,amsfonts,amsthm,amsmath}
\usepackage{graphicx,color}
\usepackage{psfrag,epsf}
\usepackage{enumitem}
\usepackage{natbib}
\usepackage[colorlinks=true, linkcolor=blue, citecolor=blue]{hyperref}
\usepackage{multirow}
\usepackage{mathrsfs}
\usepackage{bm}
\usepackage{bbm}
\usepackage{url} 
\usepackage{authblk}
\usepackage{subcaption} 
\usepackage{setspace} 

\newtheorem{lemma}{Lemma}
\newtheorem{theorem}{Theorem}
\newtheorem{proposition}{Proposition}
\newtheorem{corollary}{Corollary}
\newtheorem{assumption}{Assumption}

\newcommand{\cov}{\mbox{cov}}
\newcommand{\var}{\mbox{var}}

\newcommand{\dif}{{\rm d}}
\newcommand{\vz}{\boldsymbol{z}}

\newcommand{\reals}{\mathbb{R}}
\newcommand{\mI}{{\mathsf I}}

\newcommand{\mW}{{\mathsf W}}
\newcommand{\mZ}{{\mathsf Z}}

\newcommand{\mU}{{\mathsf U}}
\newcommand{\mV}{{\mathsf V}}
\newcommand{\cm}{\mathcal{M}}
\newcommand{\vzero}{\boldsymbol{0}}
\newcommand{\Xdes}{\mathcal{X}}

\newcommand{\mA}{{\mathsf A}}
\newcommand{\vx}{\boldsymbol{x}}
\newcommand{\vt}{\boldsymbol{t}}
\newcommand{\vX}{\boldsymbol{X}}

\newcommand{\ch}{\mathcal{H}}

\newcommand{\Tdes}{\mathcal{T}}
\newcommand{\ip}[3][{}]{\ensuremath{\left \langle #2, #3 \right \rangle_{#1}}}

\newcommand{\linearK}{K_{\text{Linear}}}

\newcommand{\norm}[2][{}]{\ensuremath{\left \lVert #2 \right \rVert}_{#1}}

\DeclareMathOperator*{\E}{\mathbb{E}}

\DeclareMathOperator*{\corr}{corr}

\title{Kernel Discrepancy-Based Rerandomization for Controlled Experiments}
\author[1]{Yiou Li}
\author[2]{Lulu Kang}
\affil[1]{DePaul University}
\affil[2]{University of Massachusetts Amherst}
\date{}
\begin{document}

\maketitle

{\singlespacing
\begin{abstract}
This paper introduces a kernel discrepancy-based framework for rerandomization to enhance the precision of causal inference in controlled experiments. 
We demonstrate that the kernel discrepancy is the key part of the variance upper bound for the difference-in-means estimator, thereby establishing a theoretical rationale for its use. 
It quantifies the difference between empirical covariate distributions of treatment groups. 
We can choose a suitable kernel function and the corresponding discrepancy to accommodate simple or complex relationships between the outcome and the covariates. 
The proposed framework efficiently applies to any number of treatment groups, overcoming a significant limitation of existing methods.
Furthermore, we develop a computationally efficient composite strategy for factorial experiments by recursively applying two- or multi-group rerandomizations. 
Numerical studies demonstrate that our approach significantly reduces estimator variance, with the linear kernel being optimal for linear relationships and the $\mathcal{L}_2$-discrepancy offering robust performance under model uncertainty.
\end{abstract}

\noindent%
}

\newpage
\singlespacing
\section{Introduction}

Controlled experiments establish causality by testing factors on experimental units. 
This process involves two steps: first, designing the treatment settings/levels (or factor-level combinations), and second, assigning these treatments to experimental (or test) units. 
While a factorial design determines the $L$ treatment settings \citep{wu2011experiments}, a completely randomized assignment of these treatment settings \citep{rubin1974estimating,rubin1980randomization} is common but cannot ensure covariate balance across different treatment groups, and thus leads to potential confounding with the treatment effects. 
This paper focuses on the assignment step and aims to improve the accuracy of the standard difference-in-means estimator for treatment effects.

Rerandomization has emerged as a powerful strategy to improve the precision of causal inference in controlled experiments by enforcing balance on observed covariates.
The foundational work by \cite{morgan2012rerandomization} formalized the concept, proposing the use of the Mahalanobis Distance (MD) as the criterion measuring covariate balance. 
In a rerandomized experiment, the test units are randomly partitioned into $L$ groups. 
The random partition is accepted if the covariate balancing criterion is met and the experiment is conducted based final partition. 
The rerandomization retains the advantages of randomization while still ensuring covariate balance.
Since this seminal work, many other works have followed the pursuit of proposing new rerandomization designs. 
\cite{morgan2015rerandomization} introduced rerandomization in tiers of covariates to improve accuracy. 
To partition the test units into tiers, different weights are assigned to covariates to represent their relative importance. 
Subsequent theoretical developments have established the asymptotic properties of the difference-in-means estimator under rerandomization, revealing a non-Gaussian sampling distribution that leads to more precise estimation and shorter confidence intervals for the average treatment effect \citep{li2018asymptotic, yang2023rejective}. 
The methodology has been extended beyond the simple treatment-control setup to more complex designs, including factorial experiments \citep{branson2016improving, zhao2018randomization}, stratified and sequential experiments \citep{zhou2018sequential, wang2023rerandomization}, and cluster-randomized trials \citep{lu2023design}.

While the MD-rerandomization is widely used, recent research has focused on developing some tailored variants. 
A key limitation of MD is that it balances all covariates equally, which can be suboptimal when covariates vary in their importance or are highly collinear. 
In response, several refined approaches have been proposed. 
These include ridge rerandomization to handle collinearity \citep{BRANSON2021287}, PCA rerandomization for high-dimensional settings \citep{zhang2024pca}, and the use of a Bayesian criterion to prioritize covariates strongly associated with the outcome \citep{liu2025bayesian}. 
Furthermore, the interplay between design and analysis has been clarified, showing that combining rerandomization with regression adjustment in the analysis stage yields further efficiency gains without sacrificing validity \citep{li2020rerandomization}. 
To address computational bottlenecks, methods like pair-switching rerandomization \citep{zhu2023pair} have been introduced, making the method more practical for applications requiring numerous randomization tests or confidence intervals.

Alternative to rerandomization, \cite{kallus2018optimal} proposed a new kernel allocation to divide the experimental units into balanced groups, which optimizes a kernel-based criterion. 
\cite{li2021covariate} introduced a new covariate balancing criterion, which measures the differences between the kernel density estimates of the covariates of the treatment groups. 
The partition of the test units is through minimizing the criterion. 
\cite{johansson2021optimal} and \cite{kallus2021optimality} discussed the controversy of the optimization-based partition, as it cannot make randomization inference. 
But this is not the focus of this paper.  

In this paper, we introduce a novel rerandomization framework that uses kernel discrepancy as a unified covariate balance criterion. 
Under the assumptions aligned with the foundational work on rerandomization, we derive the exact variance of the difference-in-means estimator and establish that the kernel discrepancy is the key component of its upper bound. 
This discrepancy measures the difference between the empirical distributions of covariates across treatment groups. 
Although it is a model-free criterion, it considers the complexity of the relationship between the experimental outcome and the covariates. 
For example, if this relationship is linear, we can choose the simple linear kernel, and thus the corresponding linear kernel discrepancy can only detect the difference in the sample means of two groups of covariates. 
In this case, the linear discrepancy is equal to MD (only for $L=2$) if the covariates are scaled and decorrelated. 
However, if we suspect this relationship is more complicated, choosing a more suitable kernel can lead to a stronger discrepancy that measures the distribution difference beyond lower-order moments.  

Besides the generality and flexibility, the kernel discrepancy's true advantage lies in its unique mathematical properties. 
Crucially, Theorem \ref{thm:discequality} reveals a property that allows it to efficiently apply to any number of treatment groups ($L \geq 2$). 
This resolves a significant gap in the literature, which has focused predominantly on two-group experiments. 
Existing multi-group extensions, which rely on pairwise criteria, face severe computational and theoretical hurdles in determining acceptance thresholds. 
Our framework overcomes these issues.

Furthermore, we leverage this property and develop a composite rerandomization strategy for large factorial experiments.
Taking the $L=2^s$ case as an example, this method partitions experimental units via a tree structure, breaking down a complex \(L=2^s\) assignment into a sequence of simple two-group rerandomizations. 
This makes it computationally feasible to achieve superior covariate balance even when $L=2^s$ is large.
Examples are shown to validate the practical utility of the proposed method. 
Rerandomization using kernel discrepancy has consistently reduced estimator variance compared to complete randomization.

\section{Preliminary: Kernel Discrepancy}\label{sec:preliminary}

We review the preliminary background on kernel discrepancy in this section.
Let $\vz$ be the $d$-dimensional covariates of any test unit and assume $\vz \in \Omega \subseteq \mathbb{R}^d$.
Let $(\cm, \ip[\cm]{\cdot}{\cdot})$ be a Hilbert space of measures defined on $\Omega$ with the inner product $\ip[\cm]{\cdot}{\cdot}$ and $\cm$ includes all Dirac measures.
Define a kernel function $K:\Omega \times \Omega \to \reals$ in terms of the inner product of Dirac measures
\begin{equation} \label{eq:kerDeltaDef}
K(\vt,\vx) := \ip[\cm]{\delta_{\vt}}{\delta_{\vx}}, \qquad \text{ for any } \vt, \vx \in \Omega.
\end{equation}
Then the squared distance between any two Dirac measures in $\cm$ is
\begin{equation} \label{eq:distDelta}
\norm[\cm]{\delta_{\vx} - \delta_{\vt}}^2 = K(\vt,\vt) - 2K(\vt,\vx) + K(\vz,\vx), \qquad \text{for any } \vt, \vx \in \Omega.
\end{equation}
It is straightforward to show that $K$ defined by \eqref{eq:kerDeltaDef} is a symmetric and positive definite kernel, namely
\begin{subequations} \label{eq:sympd}
\begin{gather}
K(\vx, \vt) = K(\vt, \vx) \qquad \text{for any } \vt, \vx\in \Omega,\\
\sum\limits_{i, k=1}^N a_i a_k  K(\vx_i,\vx_k) > 0, \qquad \text{for any } N\in\mathbb{N}, \  \bm a \in\mathbb{R}^N \setminus \{\vzero\},  \ \vx_i, \vx_j\in \Omega.
\end{gather}
\end{subequations}
So the inner product of any two measures $\lambda, \nu \in \cm$ can be expressed in terms of a double integral of the kernel $K$
\begin{equation} \label{eq:ipMdef}
\ip[\cm]{\lambda}{\nu} = \int_{\Omega \times \Omega} K(\vt,\vx) \, \lambda(\dif \vt) \nu(\dif \vx).
\end{equation}
On the other hand, a symmetric and positive definite kernel function $K:\Omega \times \Omega \to \reals$ can also define a unique Hilbert space of measures, denoted by $\cm$, whose inner product is given by \eqref{eq:ipMdef}.
It can be shown that \eqref{eq:ipMdef} is a valid inner product and $\cm$ can include all Dirac measures \citep{hickernell2016trio,li2020transformed}.

The discrepancy, which measures the difference between two probability measures $\nu$ and $\lambda$ in $\cm$, is defined as the norm of $\nu-\lambda$, i.e.,
\begin{subequations} \label{eqn:discdef}
\begin{equation} \label{eq:discDefmeasure}
D(\nu,\lambda; K)  := \norm[\cm]{\nu - \lambda} 
 = \left[\int_{\Omega \times \Omega} K(\vt,\vx) \, (\nu - \lambda)(\dif \vt) (\nu - \lambda)(\dif \vx)\right]^{1/2}.
\end{equation}
Consider two sets of observed covariates $\mathcal{Z} = \{\vz_1,\cdots,\vz_n\}$ and $\mathcal{Z}' = \{\vz'_1,\cdots,\vz'_m\}$ with empirical distribution functions (or empirical CDF) $F_{\mathcal{Z}}$ and $F_{\mathcal{Z}'}$, respectively.
Corresponding to $F_{\mathcal{Z}}$ and $F_{\mathcal{Z}'}$, we can also define two measures $\nu_{\mathcal{Z}}$ and $\nu_{\mathcal{Z}'}$. 
It is obvious that $\nu_{\mathcal{Z}}, \nu_{\mathcal{Z'}}\in \cm$.
The discrepancy between $F_{\mathcal{Z}}$ and $F_{\mathcal{Z}'}$, or equivalently $\nu_{\mathcal{Z}}$ and $\nu_{\mathcal{Z}'}$,  is
\begin{align}\nonumber
& D\left(F_{\mathcal{Z}},F_{\mathcal{Z}'};K\right) = D(\nu_{\mathcal{Z}},\nu_{\mathcal{Z}'}; K) \\\label{eq:discDefdiscrete}
= & \left[\frac{1}{m^2}  \sum_{i,k=1}^{m^2} K(\vz'_i,\vz'_k) - \frac 2{nm} \sum_{i,k=1}^{mn}  K(\vz_i,\vz'_k)  + \frac{1}{n^2}  \sum_{i,k=1}^{n^2} K(\vz_i,\vz_k)\right]^{1/2}. 
\end{align}
\end{subequations}

Besides being considered as a norm on the Hilbert space of measures, \eqref{eq:discDefmeasure} can also be interpreted as a deterministic quadrature error bound for the Monte Carlo method as explained in \cite{hickernell2016trio} and \cite{li2020transformed}.
Specifically, a symmetric and positive definite function $K$ defines a unique Hilbert space $\mathcal{H}: \Omega\times\Omega \rightarrow \reals$ of functions $f: \Omega \rightarrow \reals$ with an inner product $\ip[\ch]{\cdot}{\cdot}$. 
The Hilbert space $\ch$ and the inner product satisfy the following conditions: $K(\cdot, \bm z)\in\mathcal{H}$, $\ip[\ch]{K(\cdot, \vz)}{K(\cdot, \vx)}=K(\vx,\vz)$, and $f(\vz)=\ip[\ch]{f}{K(\cdot,\vz)}$ for any $\vz, \vx \in\Omega$ and any $f\in\ch$.
The Hilbert space $\mathcal{H}$ is called \emph{reproducing kernel Hilbert space} (RKHS) defined by the \emph{reproducing kernel} $K$ \citep{aronszajn1950theory,berlinet2011reproducing}.
For any $f\in\mathcal{H}$, the following inequality holds \citep{hickernell1998generalized}
\begin{equation}\label{eqn:integralbound}
\left|\int_{\Omega} f(\vz)d F_{\mathcal{Z}}(\vz) - \int_{\Omega} f(\vz)d F_{\mathcal{Z}'}(\vz)\right|\leq D\left(F_{\mathcal{Z}},F_{\mathcal{Z}'};K\right)V(f),
\end{equation}
where $D\left(F_{\mathcal{Z}},F_{\mathcal{Z}'};K\right)$ is the discrepancy in \eqref{eq:discDefdiscrete} and $V(f)$ is the \emph{variation} of any function $f\in\mathcal{H}$, which reflects the smoothness of the function $f$.
The variation $V(f)$ has two different formats depending on whether the constant function $f(\bm x)=c$ is in $\ch$.
\begin{equation} \label{eqn:variation}
V(f)=\begin{cases} \|f\|_{\mathcal{H}}&\text{if} \,\,1\notin\mathcal{H},\\
\left(\|f\|^2_{\mathcal{H}}-\langle
f,1\rangle^2_{\mathcal{H}}/\|1\|^2_{\mathcal{H}}\right)^{1/2}&\text{if}\,\,
1\in\mathcal{H}.\end{cases}
\end{equation}
The equality in \eqref{eqn:integralbound} is attainable for at least one $f\in \mathcal{H}$.
In fact, \eqref{eqn:integralbound} is a generalization of the Koksma-Hlawka inequality \citep{brandolini2013koksma} because in \eqref{eqn:integralbound} both $F_{\mathcal{Z}}$ and $F_{\mathcal{Z}'}$ are empirical measures of finite sets of samples, whereas the original Koksma-Hlawka inequality involves one continuous and one empirical measure.
The kernel discrepancy defined in \eqref{eq:discDefmeasure}, its special case \eqref{eq:discDefdiscrete}, and the inequality \eqref{eqn:integralbound} provide the theoretical foundation for using kernel discrepancy as a rerandomization criterion.

In the machine learning community, the kernel discrepancy is better-known as the \emph{Maximum Mean Discrepancy} or \emph{MMD}. 
It is widely applied in various topics such as generative models \citep{arbel2019maximum, pmlr-v37-li15}, Bayesian statistics \citep{pmlr-v118-cherief-abdellatif20a}, robust learning \citep{alquier2024universal}, etc. 
Particularly, \cite{gretton2012kernel} introduced a two-sample test based on MMD. 
Although sharing similar basic intuition about using MMD to measure the difference of distributions, our theories and methodologies are under the assumption of finite samples, which is entirely different from these works. 

\section{Difference-in-means Estimator}
\label{sec:framework}

In this section, we set up the notation for controlled experiments and the necessary assumptions regarding the underlying model and rerandomization.
Based on this general framework, we discuss the bias and variance properties of the widely used difference-in-means estimator.

Consider an $L$-level experiment with $N$ test units, where $L$ is the number of treatment settings and $L\geq 2$.
A $d-$dimensional covariates $\vz = [z_1,...z_d]^\top\in \Omega$ is observed for a test unit.
Denote $\vz_i = [z_{i1},...,z_{id}]^\top$ as the covariates of the $i$-th test unit and $\mZ = (\bm z_1,...,\bm z_N)^\top$ is the $N\times d$ matrix of covariates of the $N$ test units.
In this paper, we assume the $N$ test units are pre-selected from the target population of the experimental study.
Consequently, $\mZ$ is known before the experiment.
Once the $L$ treatment settings of the involved experimental factors are chosen (for example, a $2^{p-k}=L$ factorial experiment design), the remaining task is to assign a treatment setting to each test unit.
The focus of this paper is on how to assign a treatment setting to each test unit given the covariates matrix $\mZ$.
Let $\vX = [X_1,\ldots, X_N]^\top$ be the assignment vector of length $N$, where $X_i$ is the treatment level assigned to the $i$-th test unit.
The $L$ treatment levels are labeled by $\{1,\ldots, L\}$ and thus $X_i\in \{1,\ldots, L\}$.
The following Assumption \ref{ass:model} is a common model assumption for controlled experiments, which states that the effects of the treatment factors and covariates are additive. 

\begin{assumption}\label{ass:model}
The response of the $i$-th test unit when assigned to treatment level $q$ is
\begin{equation}\label{model:rm}
Y_i(q) =\alpha_q+f(\vz_i)+\epsilon_i,\quad q = 1,\ldots,L,
\end{equation}
where $\alpha_q$ is the treatment effect of level $q$ and the function $f:\Omega \rightarrow \reals$ is the effect of covariates to the response.
The random noise $\epsilon_i$'s are iid following a certain distribution with zero mean and constant variance $\sigma^2$, and are also independent of the treatment assignments and covariates of all test units.
\end{assumption}

Using the potential outcome framework \citep{rubin2005causal}, the observed response of the $i$-th test unit is denoted by $Y_{obs,i}$ and
\begin{equation}\label{model:obsY}
Y_{obs,i}=\sum_{l=1}^L Y_i(l)\mathbbm{1}(X_i=l),
\end{equation}
where $\mathbbm{1}(\cdot)$ is the indicator function.
In practice, the targets of inference are the differences between two treatment effects, $\alpha_q-\alpha_{q'}$ for $q,q'=1,\ldots,L$.
A common estimator for the contrast $\alpha_q-\alpha_{q'}$ is the difference-in-means estimator
\begin{equation}\label{eqn:difference-in-means}
\hat{\alpha}_{q}-\hat{\alpha}_{q'}=\bar{Y}_{obs}(q)-\bar{Y}_{obs}(q'), \quad q,q'=1,\ldots,L,
\end{equation}
where 
\[\bar{Y}_{obs}(q) = \frac{\sum_{i=1}^N Y_{obs,i}\mathbbm{1}(X_i=q)}{\sum_{i=1}^N\mathbbm{1}(X_i=q)}, \quad \text{for }q\in\{1,\ldots, L\}.\]
It is the sample means of the responses of treatment group $q$. 
Based on \eqref{model:obsY}, $\bar{Y}_{obs}(q)$ can also be expressed by
\[
\bar{Y}_{obs}(q)=\frac{\sum_{i=1}^N Y_i(q)\mathbbm{1}(X_i=q)}{\sum_{i=1}^N \mathbbm{1}(X_i=q)}.
\]

We first answer the question: Can rerandomization improve the accuracy of the difference-in-means estimator?
Following the same notation in \cite{morgan2012rerandomization}, given a pre-specified rerandomization criterion, we use $\phi(\mZ,\vX)$ to indicate whether the rerandomization criterion is met for the treatment assignments $\vX$, i.e., 
\begin{equation}\label{eqn:gen_rerandcri}
\phi(\mZ,\vX) = \left\{\begin{array}{ll}
1,\,\,\,&\text{if $\vX$ is an acceptable randomization;}\\
0,\,\,\,&\text{if $\vX$ is not an acceptable randomization.}\\
\end{array}\right.
\end{equation}
To facilitate the later discussion, we make the following assumptions on the randomized assignment and the rerandomization condition. 

\begin{assumption}\label{ass:design} The randomized assignment $\bm X=[X_1,\ldots, X_N]$ and rerandomization condition $\phi(\mZ, \bm X)$ satisfy the following conditions. 
\begin{enumerate}
\item If $\phi(\mZ,\bm X)=1$, conditional on the covariates matrix $\mZ$, the random treatment assignment $X$ of any test unit does not depend on any response $Y$. 
\item The sample size $N$ is fixed prior to the experiment and $N$ is divisible by $L$. The random $\bm X$ satisfying $\phi(\mZ, \bm X)=1$ splits the $N$ test units into $L$ groups of the same size, i.e., $\sum_{i=1}^N \mathbbm{1}(X=l)=N/L=n$ for any $l\in \{1,\ldots,L\}$. 
\item Let $\pi$ be any permutation of $1,\ldots, L$. The rerandomization criterion satisfies $\phi(\mZ, \bm X)=\phi(\mZ, \pi(\bm X))$. 
\end{enumerate}
\end{assumption}

The first two conditions of Assumption \ref{ass:design} are common for controlled experiments. 
For any complete randomization, they are true without considering $\mZ$ or $\phi(\mZ, \bm X)=1$. 
But for a rerandomized assignment with pre-selected $N$ test units, the first two conditions do not hold automatically and thus need to be explicitly stated. 
Also, the equal sample size in each treatment group leads to $\Pr(X_i=q)=\E(\mathbbm{1}(X_i=q))=1/L$ and $\Pr(X_i=q|\mZ, \phi(\mZ, \bm X)=1)=\E(\mathbbm{1}(X_i=q)|\mZ, \phi(\mZ, \bm X)=1)=1/L$. 
The last condition simply means that changing the labeling order of the $L$ treatments does not change the rerandomization criterion. 
For $L=2$, this can be simply written as $\phi(\mZ, \bm X)=\phi(\mZ, {\bf 1}-\bm X)$, which is the same assumption used in Theorem 2.1 in \cite{morgan2012rerandomization}.  
Based on Assumptions \ref{ass:model} and \ref{ass:design}, Proposition \ref{prop:mean-var} gives the mean and variance of the difference-in-means estimator of the rerandomized assignment. 
Note that expectation is with respect to both the response variable and the random assignment.
Proofs and derivations are in the Supplementary Material. 

\begin{proposition}\label{prop:mean-var}
Given the covariates matrix $\mZ$ of the $N$ test units and a pre-specified rerandomization criterion, under Assumption \ref{ass:model} and \ref{ass:design}, the difference-in-means estimator $\hat{\alpha}_{q}-\hat{\alpha}_{q'}$ is an unbiased estimator of $\alpha_q-\alpha_{q'}$ for any $q, q'=1,\ldots, L$, i.e.,
\[\E[\hat{\alpha}_{q}-\hat{\alpha}_{q'}|\mZ,\phi(\mZ,\vX)=1]=\alpha_q-\alpha_{q'}, \quad q,q'=1, \ldots, N,\]
and the variance of the difference-in-means estimator is
\begin{align}\nonumber
&\var[\hat{\alpha}_{q}-\hat{\alpha}_{q'}|\mZ,\phi(\mZ,\vX)=1]\\\label{eqn:variance}
=&\E\left[\left.\left(\int f(\vz)\dif F_{q}(\vz)-\int f(\vz)\dif F_{q'}(\vz)\right)^2\right|\mZ,\phi(\mZ,\vX)=1\right]+ \frac{2L}{N}\sigma^2,
\end{align}
where $F_q$ and $F_{q'}$ are empirical distribution functions of the covariates in treatment groups $q$ and $q'$, respectively.
\end{proposition}

In practice, the specific form of $f$ is usually unknown before the experiment. 
The commonly used assumptions on $f$ include a linear combination of covariates or polynomial basis functions of the covariates. 
But sometimes a more complicated functional assumption of $f$ is needed. 
To account for the uncertainty of the specific form of $f$, we assume $f$ is in an RKHS $\ch$ defined by the reproducing kernel $K$. 
We should choose $K$ such that the corresponding $\ch$ contains all potential forms of $f$ for the experiment. 
Based on the property of $\ch$, we can find an upper bound for the variance of the difference-in-means estimator for any $f$ in $\ch$. 

\begin{theorem}\label{thm:mse}
Suppose that $\mathcal{H}$ is the RKHS of functions defined on $\Omega$ with the reproducing kernel $K$, such that the $f$ in \eqref{model:rm} is in $\ch$. 
Under both Assumption \ref{ass:model} and \ref{ass:design}, the variance of the difference-in-means estimator $\hat{\alpha}_{q}-\hat{\alpha}_{q'}$ has an upper bound given by
\begin{equation}\label{ineqn:discrepancy}
\var[\hat{\alpha}_{q}-\hat{\alpha}_{q'}|\mZ,\phi(\mZ,\vX)=1]\leq \E\left[\left.D^2\left(F_q,F_{q'};K\right)\right|\mZ,\phi(\mZ,\vX)=1\right]\left[V\left(f\right)\right]^2+\frac{2\sigma^2}{n},
\end{equation}
where $D(F_q,F_{q'};K)$ is the discrepancy between the empirical distribution of covariates in treatment group $q$ and $q'$ defined in \eqref{eq:discDefdiscrete} and $V\left(f\right)$ is the variation of function $f$ defined in \eqref{eqn:variation} that does not depend on the treatment assignment $\vX$.
In the upper bound, the expectation is with respect to $\bm X$. 
\end{theorem}

Here are some remarks regarding Theorem \ref{thm:mse}.
First, $\E\left[\left.D^2\left(F_q,F_{q'};K\right)\right|\mZ,\phi(\mZ,\vX)=1\right]$ is the mean squared discrepancy over all acceptable $\bm X$ as long as $\phi(\mZ, \bm X)=1$. 
This upper bound holds for any rerandomization criterion. 
If the criterion is such that $\phi=1$ for all $\vX$, then the upper bound holds complete randomization. 
Second, the upper bound \eqref{ineqn:discrepancy} separates the uncertainty of $f$ and the randomness of $\vX$. 
The mean squared discrepancy $\E\left[\left.D^2\left(F_q,F_{q'};K\right)\right|\mZ,\phi(\mZ,\vX)=1\right]$ only depends on the rerandomization criterion and the reproducing kernel $K$, but not $f$.
The variation $V(f)$ measures the roughness of $f$ and does not depend on $\vX$. 
The term $2\sigma^2/n$ is a constant too. 
To keep the upper bound of the variance small, we should choose the rerandomization criterion such that the squared discrepancy $D^2(F_q,F_{q'}; K)$ is small if $\phi(\mZ, \bm X)=1$. 
When $f$ is believed to be oscillating, that is, $V(f)$ is large, assuring a small $D^2(F_q,F_{q'}; K)$ becomes even more critical.
Theorem \ref{thm:mse} provides a natural rerandomization criterion to regulate the variance of the difference-in-means estimator. 
The equality in \eqref{ineqn:discrepancy} is attainable if there exists a function $f^*\in\mathcal{H}$ such that the equality of \eqref{eq:singleupper} (in the proof of Theorem \ref{thm:mse}) holds, i.e.,
\[\left(\int f^*(\vz)\dif F_{q}(\vz)-\int f^*(\vz)\dif F_{q'}(\vz)\right)^2 =  D^2\left(F_q,F_{q'};K\right)[V(f^*)]^2,\]
for any $\vX$ that satisfies $\phi(\mZ,\vX)=1$. 
This is more likely to occur if there is only one possible acceptable random assignment $\vX$, i.e., the optimal assignment case. 

\section{Rerandomization based on Discrepancy}\label{sec:rerandom}

An assignment $\vX$ partitions the $N$ test units into $L$ groups, and each group can be randomly assigned to one treatment level. 
Therefore, any realization of $\bm X$ corresponds to a set of empirical distributions of the covariates $F_l(\bm z)$ for $l=1,\ldots, L$. 
When $L=2$, we only need to consider the discrepancy $D^2(F_1, F_2; K)$ between the two treatment groups in the experiment. 
Naturally, we can define the rerandomization criterion as $\phi(\mZ, \bm X)=1$ if $D(F_1,F_2;K)$ is smaller than a pre-specified critical value and $\phi(\mZ, \bm X)=0$ otherwise. 
When $L\geq 3$, there are $\binom{L}{2}$ discrepancies between all pairs of treatment groups. 
An intuitive rerandomization criterion can be based on the sum of the $\binom{L}{2}$ discrepancies, that is,
\[\sum_{q,q'=1, q<q'}^L D^2\left(F_q,F_{q'};K\right).\]
It regulates the sum of the variances of all the pair-wise difference-in-means estimators. 
However, the computation of the $\binom{L}{2}$ pairs of discrepancy grows exponentially as $L$ increases. 
Interestingly, the following Theorem \ref{thm:discequality} reveals that $\sum_{q,q'=1, q<q'}^L D^2\left(F_q,F_{q'};K\right)$ is equal to $L \sum_{q=1}^L D^2(F_q,F;K)$. 
Here, for an assignment $\vX$, $F_q$ is the empirical CDF of the covariates in treatment group $q$, whereas $F$ is the empirical CDF of the covariates of all the $N$ test units. 
Consequently, we only need to compute $L$ discrepancies to check $\phi(\mZ, \bm X)$. 

\begin{theorem}\label{thm:discequality}
For any symmetric and positive-definite kernel function $K$, if $\sum_{i=1}^N \mathbbm{1}(X_i=q)=N/L=n$ for $l=1,\ldots,L$, then
\begin{equation}\label{eqn:discrepancyrelation}
\sum_{q=1}^L D^2(F_q,F;K) = \frac{1}{L}\sum_{q,q'=1,q<q'}^L D^2\left(F_q,F_{q'};K\right). 
\end{equation}
\end{theorem}

\begin{corollary}\label{cor:discequality}
 When $L=2$, for any symmetric and positive-definite kernel $K$, if $\sum_{i=1}^N \mathbbm{1}(X_i=1)=\sum_{i=1}^N \mathbbm{1}(X_i=2)=N/2=n$, then
\[D^2(F_1,F;K) = D^2(F_2,F;K) = \frac{1}{4}D^2(F_1,F_2;K).\]
\end{corollary}

We formally introduce the following discrepancy-based rerandomization criterion for any assignment $\bm X$ and given covariates $\mZ$,
\begin{align}\label{eqn:rerandcri}
\phi(\mZ,\vX) &= \left\{\begin{array}{ll}
1,\,\,\,& \sum_{q=1}^L D^2(F_q,F;K)\leq  a_p;\\
0,\,\,\,&\text{otherwise.}\\
\end{array}\right.
\end{align}
For $L=2$, due to Corollary \ref{cor:discequality}, we only need to compute one of the three possible discrepancies. 
The critical value $a_{p}$ is chosen to be the $p\times 100\%$th percentile of the distribution of $\sum_{q=1}^L D^2(F_q,F;K)$ for all random assignments. 
We choose $p$ to be a small value, which leads to a small critical value $a_p$. 
For simple kernel functions, the theoretical distribution of $\sum_{q=1}^L D^2(F_q,F;K)$ can be derived, but not so for complicated kernels. 
Details are discussed in Section \ref{sec:asymptotic}.
Theorem \ref{thm:overallbound} gives an upper bound of the sum of the variances of the difference-in-means estimators based on the proposed $\phi(\mZ, \bm X)$. 

\begin{theorem}\label{thm:overallbound}
For a controlled experiment with $L$ treatment settings and $N$ test units, under Assumption \ref{ass:model} and \ref{ass:design}, using the rerandomization criterion \eqref{eqn:rerandcri}, the sum of the variances of all the difference-in-means estimators is upper bounded by
\begin{equation}\label{ineqn:varreranddisc}
\sum_{q<q',q,q'\in\{1,\ldots,L\}}\var[\hat{\alpha}_{q}-\hat{\alpha}_{q'}|\mZ,\phi(\mZ,\vX)=1]\leq a_{p}L\left[V\left(f\right)\right]^2+\frac{\sigma^2L(L-1)}{n}.
\end{equation}
\end{theorem}

\section{Different Discrepancies}\label{sec:asymptotic}

The critical value $a_p$ depends on the probabilistic distribution of $\sum_{q=1}^L D^2(F_q, F; K)$.
Besides $N$ and $L$, it is decided by the kernel function. 
In this section, we first derive the exact distribution of $\sum_{q=1}^L D^2(F_q, F; K)$ for finite samples using a linear kernel function.  
But for more complicated kernels, we can only obtain the distribution via simulations. 

\subsection{Linear and Polynomial Discrepancy}\label{subsec:linear}
Sometimes it is evident that the response is linearly dependent on the covariates. 
Therefore, the RKHS $\mathcal{H}$ should include all linear functions of the covariates, and the corresponding reproducing kernel $K$ is the linear kernel,
\begin{equation}\label{eqn:lin_kernel}
\linearK(\vx,\vt) = 1+\vx^{\top}\vt = 1+\sum\limits_{i=1}^d x_it_i,\,\,\,\,i=1,\cdots,d. 
\end{equation}

We need to introduce some additional notation. 
Denote $z_{ik}$ as the $k$-th covariate value of the $i$-th unit, that is, the $(i,k)$ element of the covariate matrix $\mZ$. 
Denote the columns of $\mZ$ as $\bm Z_1,...,\bm Z_d$, the mean of the $k$-th covariate in the $q$-th group as 
\[\bar{Z}_k^{(q)} = \frac{\sum_{i=1}^N z_{ik}\mathbbm{1}(X_i=q)}{\sum_{i=1}^N \mathbbm{1}(X_i=q)}\]
for $q=1,...L, k=1,\ldots,d$, and the mean of the $k$-th covariate of all $N$ test units as $\bar{Z}_k$ for $k=1,...,d$. 
Without loss of generality, we assume that the finite population variances of covariates have been standardized, i.e., $\nu_k=\frac{1}{N-1}\sum_{i=1}^N(z_{ij}-\bar{Z}_k)^2 = 1$ for all $k$'s. 

\begin{proposition}\label{prop:discrelation}
For any assignment $\bm X$, $F_q$ for $q=1,\ldots, L$ is the empirical CDF of the covariates of the test units with $\mathbbm{1}(X_i=q)=1$ and $F$ is the empirical CDF of the covariates of all $N$ test units. 
With linear kernel function $K$ defined in \eqref{eqn:lin_kernel}, the squared discrepancy between the empirical distributions of the covariates in group $q$ and $q'$ can be written as
\[D^2(F_q,F_{q'};\linearK) =  \sum_{k=1}^d\left(\bar{Z}_k^{(q)}-\bar{Z}_k^{(q')}\right)^2,\]
and the squared discrepancy between group $q$ and all the $N$ test units is 
\[D^2(F_q,F;\linearK) =  \sum_{k=1}^d\left(\bar{Z}_k^{(q)}-\bar{Z}_k\right)^2.\]
\end{proposition}
Although Proposition \ref{prop:discrelation} does not require the assignment $\bm X$ to be balanced, i.e., the second condition in Assumption \ref{ass:design}, this condition is necessary for the remaining section. 
Next, we derive the asymptotic distribution of $D^2(F_1,F;\linearK)$ for a random and balanced assignment for $L=2$. 
Following Corollary \ref{cor:discequality}, we simultaneously obtain the asymptotic distribution of $D^2(F_2,F;\linearK)$ and $D^2(F_1,F_2;\linearK)$. 
Then, we extend the results to the general case of $L\geq 2$.

\begin{theorem}\label{thm:discdistrfinite}
Assume that the variances of the finite covariates are standardized, i.e., 
\[\nu_j=\frac{1}{N-1}\sum_{i=1}^N(z_{ij}-\bar{Z}_j)^2 = 1,\]
for $j=1,\cdots,d$.
Consider the random and balanced assignment $\bm X$ for $L=2$ treatment levels such that $\sum_{i=1}^N \mathbbm{1}(X_i=1)=\sum_{i=1}^N \mathbbm{1}(X_i=2)=n=N/2$. 
Then, $N\times D^2\left(F_1,F;\linearK\right)$ has the following asymptotic cumulative distribution function. 
As $N\rightarrow \infty$, for any $y>0$, 
\begin{align}\label{eqn:cumuL2}\nonumber
F_{ND^2}(y) & =\Pr\left(N\times D^2\left(F_1,F;\linearK\right)\leq y \right) \\
& = \frac{y^{d/2}}{\det(\mA^{1/2})\Gamma(1+d/2)}\times \Phi_2^{(d)}\left(\frac{1}{2},...,\frac{1}{2};1+\frac{d}{2};-\frac{y}{\lambda_1},...,-\frac{y}{\lambda_d}\right),
\end{align}
where $\Phi_2^{(d)}$ is the confluent Lauricella function, $\{\lambda_i\}_{i=1}^d$ are the eigenvalues of the $d\times d$ positive definite matrix 
\[\mA = 2\begin{pmatrix}
1 & |\rho_{12}| & \cdots & |\rho_{1d}|\\
|\rho_{21}| & 1 & \cdots & |\rho_{2d}|\\
\vdots & \vdots & \ddots & \vdots\\
|\rho_{d1}| & \cdots & \cdots & 1
\end{pmatrix},
\]
and $\rho_{ik}$ is the correlation between columns $\bm Z_j$ and $\bm Z_k$. 
\end{theorem}

Evaluating the confluent Lauricella function $\Phi_2^{(d)}$ in \eqref{eqn:cumuL2} is computationally expensive and not available in most software. 
Fortunately, \cite{ferrari2019note} provides an approximation to the distribution of the summed correlated Chi-square random variables. 
Using this result, we obtain two different Gamma distributions to approximate the asymptotic distributions of $N\times D^2(F_1, F; \linearK)$ and $N\times D^2(F_1,F_2;\linearK)$. 
For the $\text{Gamma}(a,b)$ distribution, $a$ is the shape parameter and $b$ is the scale parameter. 

\begin{theorem}\label{thm:discfiniteapprox}
Under the same assumptions of Theorem \ref{thm:discdistrfinite}, approximately, as $N\rightarrow\infty$,
$$N \times D^2\left(F_1,F;\linearK\right) \stackrel{\text{d}}{\rightarrow}\text{Gamma}\left(\frac{d}{u}, u \right),$$
and
$$N \times D^2\left(F_1,F_2;\linearK\right) \stackrel{\text{d}}{\rightarrow}\text{Gamma}\left(\frac{d}{u}, 4u\right),$$
where $u = 2\left(1+\frac{2}{d} \sum\limits_{j\neq k}^d\rho^2_{jk}\right)$. 
\end{theorem}

Since $D^2\left(F_1,F;\linearK\right)=D^2\left(F_2,F;\linearK\right)$ for any random and balanced assignment $\bm X$, $N\times D^2\left(F_2,F;\linearK\right)$ has the same asymptotic distribution and the approximated asymptotic distribution of $N\times D^2\left(F_1,F;\linearK\right)$. 
Theorem \ref{thm:linearL3approx} extends the results to the general case of $L\geq 2$. 
It is clear that when $L=2$, the result in Theorem \ref{thm:linearL3approx} renders to the result in Theorem \ref{thm:discfiniteapprox}. 

\begin{theorem}\label{thm:linearL3approx}
Assume that the finite covariates are standardized, i.e., 
\[\nu_j=\frac{1}{N-1}\sum_{i=1}^N(z_{ij}-\bar{Z}_j)^2 = 1\]
for $j=1,\cdots,d$. 
Consider the random and balanced assignment $\bm X$ for $L\geq 2$ treatment levels such that $\sum_{i=1}^N \mathbbm{1}(X_i=q)=n=N/L$ for $q=1,\ldots, L$. 
As $N\rightarrow \infty$, approximately, 
\[
N\times \sum_{q=1}^L D^2(F_q,F;\linearK) \stackrel{\text{d}}{\rightarrow} \text{Gamma}\left(\frac{(L-1)d}{u},Lu\right),
\]
where $u = 2\left(1+\frac{2}{d}\sum\limits_{j\neq k}^d\rho^2_{jk}\right)$ and $\rho_{jk}$ is the correlation between columns $\bm Z_j$ and $\bm Z_k$. 
\end{theorem}

Based on Theorem \ref{thm:linearL3approx}, for any medium or large $N$, approximately, 
\[
\sum_{q=1}^L D^2(F_q, F; \linearK) \sim \text{Gamma}\left(\frac{(L-1)d}{u},\frac{Lu}{N}\right). 
\]
When the linear kernel in \eqref{eqn:lin_kernel} is used, we can choose $a_p$ as the $p\times 100\%$th percentile from this Gamma distribution, and discrepancy-based rerandomization criterion $\phi(\mZ, \bm X)=1$ if $\sum_{q=1}^L D^2(F_q, F; \linearK)\leq a_p$.

In the following two examples, we use simulated covariates to show that the Gamma distribution is almost identical to the empirical distribution of $\sum_{q=1}^L D^2(F_q, F; \linearK)$ obtained from 1000 randomly generated assignments satisfying the assumptions of Theorem \ref{thm:linearL3approx}. 

\begin{figure}[htb]
\centering
\begin{subfigure}{.45\linewidth}
\includegraphics[width=\linewidth]{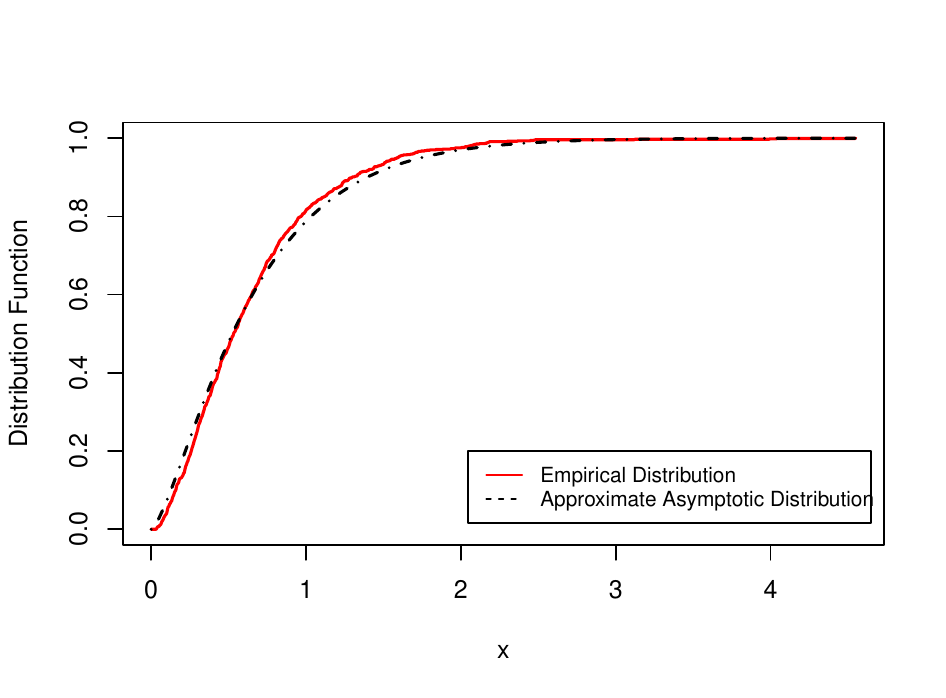}
\caption{$N=30$}
\end{subfigure}
\begin{subfigure}{.45\linewidth}
\includegraphics[width=\linewidth]{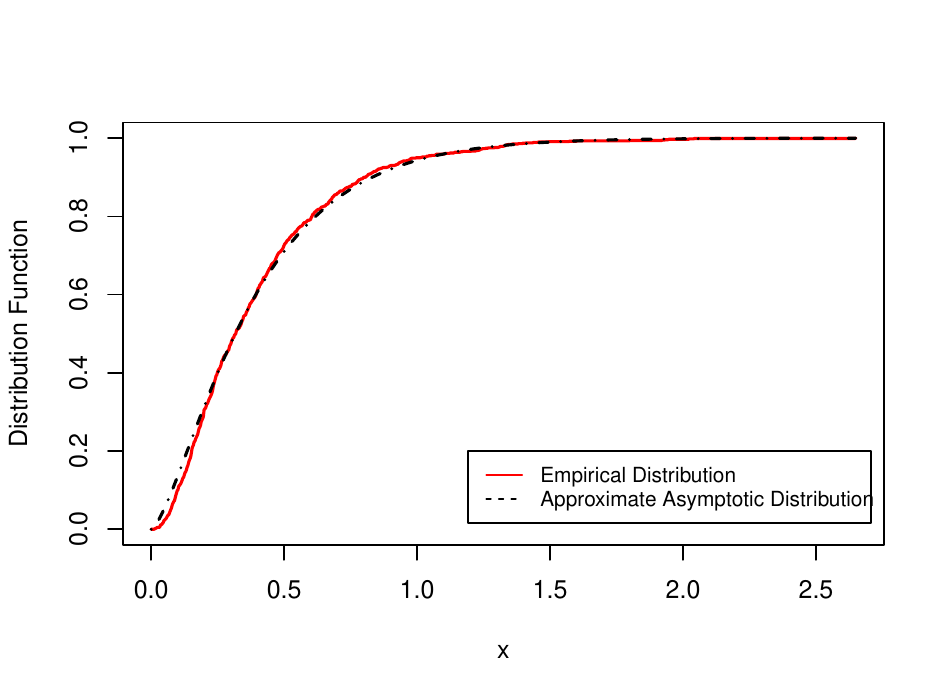}
\caption{$N=50$}
\end{subfigure}
\begin{subfigure}{.45\linewidth}
\includegraphics[width=\linewidth]{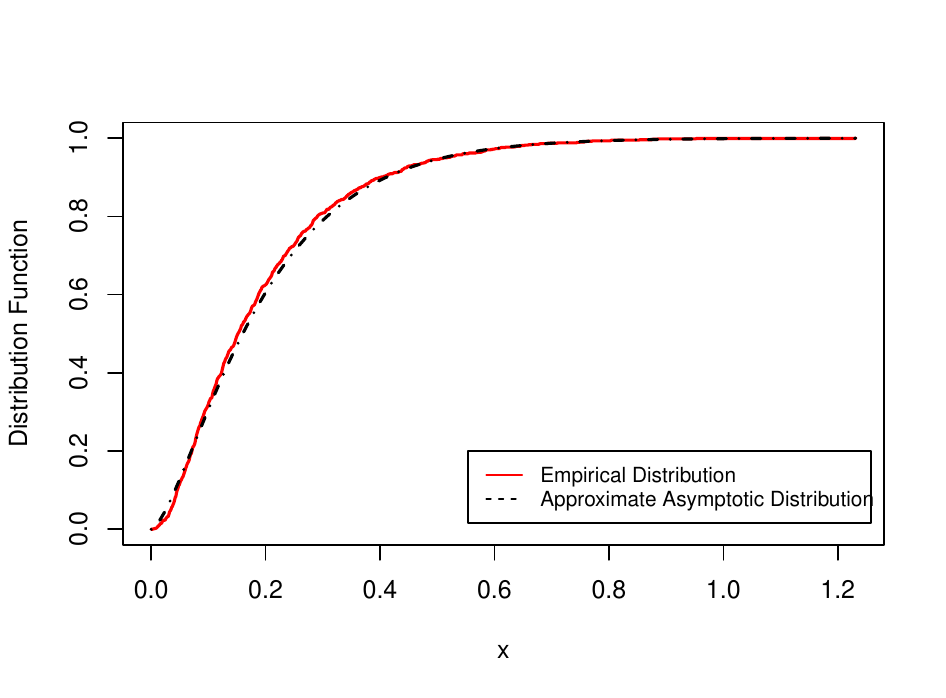}
\caption{$N=100$}
\end{subfigure}
\begin{subfigure}{.45\linewidth}
\includegraphics[width=\linewidth]{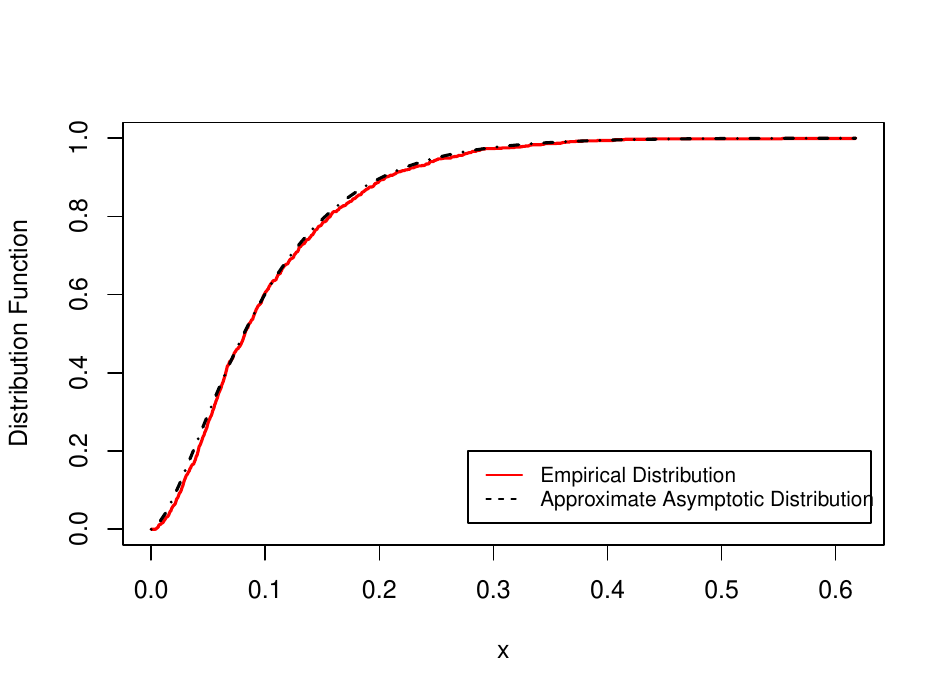}
\caption{$N=200$}
\end{subfigure}
\caption{Empirical Distribution of $D^2\left(F_1,F_2;K\right)$ (red solid curve) v.s. the approximate Gamma distribution (black dashed curve).}
\label{fig:AsympDistL2}
\end{figure}

\noindent{\bf Example 1}.  We generate the covariate matrix $\mZ = (\bm Z_1,...,\bm Z_d)$ for $d=5$ and different $N$ as follows. 
For the first two of columns $\bm Z_1$ and $\bm Z_2$, we generate $70\%$ of the samples from $N(\bm\mu_1,\Sigma)$, and $30\%$ from $N(\bm\mu_2,\Sigma)$, where $\bm\mu_1=-3\times{\bf 1}$,  $\bm\mu_2=5\times {\bf 1}$, with ${\bf 1}$ as the vector of 1's. 
The matrix $\Sigma$ is a random positive definite matrix.
For $\bm Z_3$, we generate $40\%$ of the samples from Uniform$(-0.5,1.5)$ and $60\%$ from Uniform$(-3,8)$.
For $\bm Z_4$, we generate $20\%$ of the samples from Gamma$(0.1,1)$ and $80\%$ from Gamma$(2.5,1)$.
For $\bm Z_5$, we generate $30\%$ of the samples from $N(0.05,1)$ and $70\%$ from $N(10,1)$.
Figure \ref{fig:AsympDistL2} compares the empirical distribution (y-axis) of squared discrepancy $D^2\left(F_1,F_2;\linearK\right)$ (x-axis) of 1000 random and balanced assignments and the approximate distribution $\text{Gamma}\left(\frac{d}{u}, \frac{4u}{N}\right)$ in Theorem \ref{thm:discfiniteapprox} for $N=30, 50, 100,$ and 200.

\begin{figure}[htb]
\centering
\begin{subfigure}{.45\linewidth}
\includegraphics[width=\linewidth]{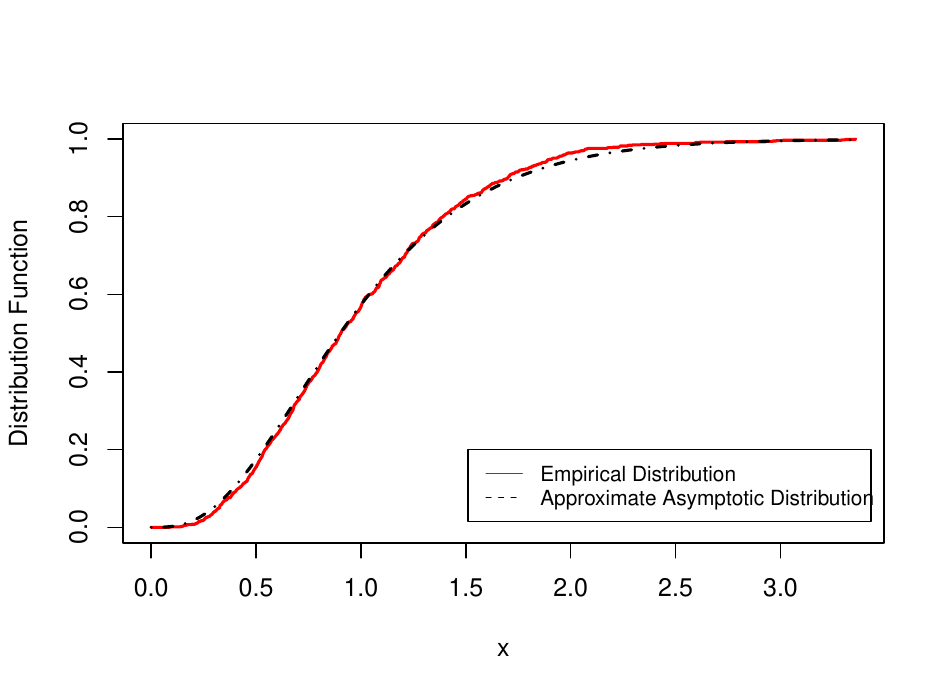}
\caption{$N=30$}
\end{subfigure}
\begin{subfigure}{.45\linewidth}
\includegraphics[width=\linewidth]{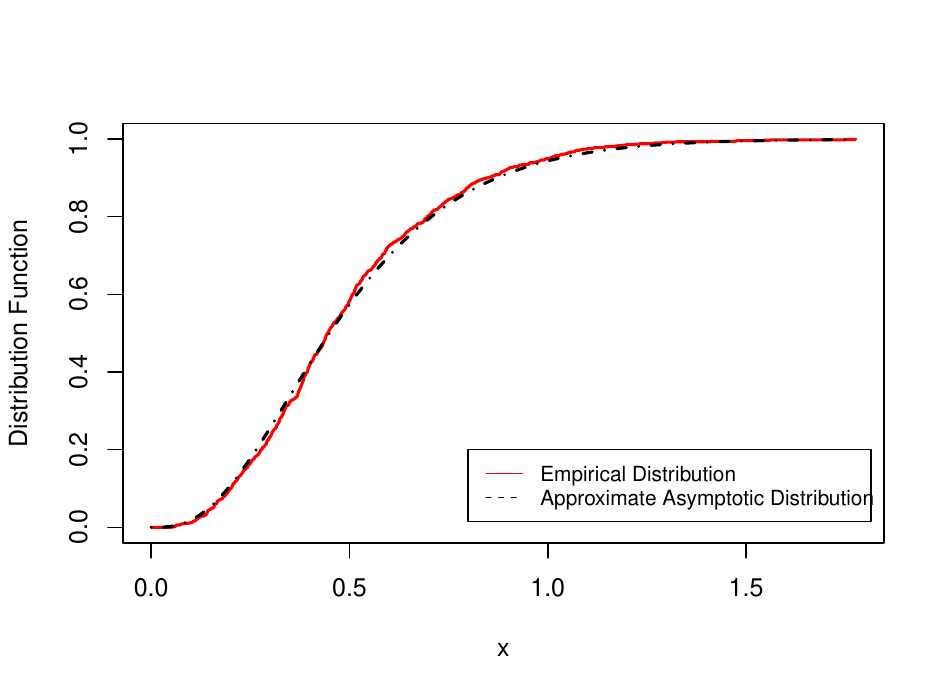}
\caption{$N=60$}
\end{subfigure}
\begin{subfigure}{.45\linewidth}
\includegraphics[width=\linewidth]{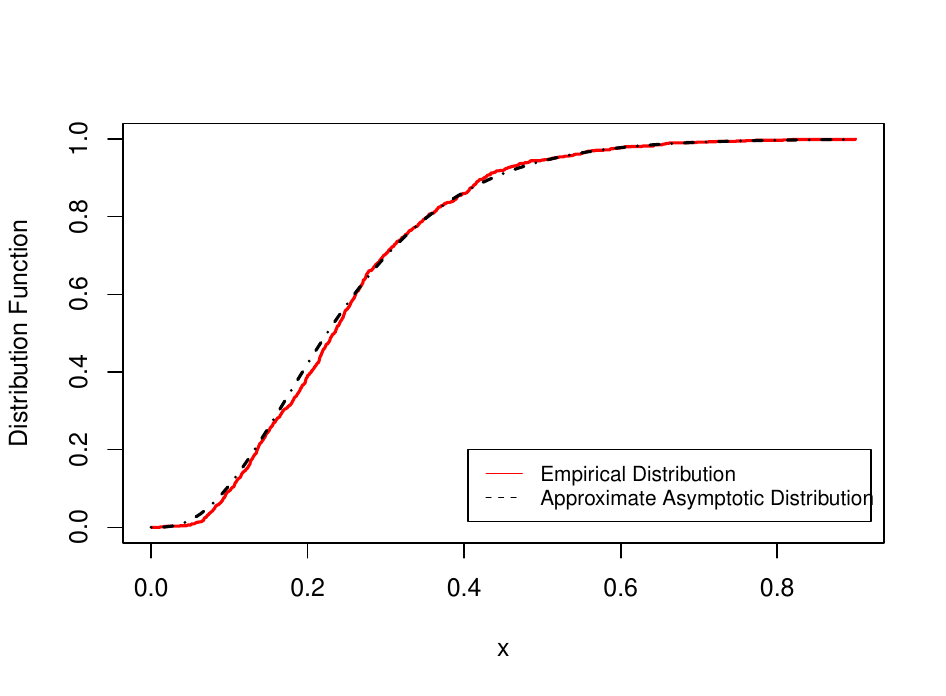}
\caption{$N=120$}
\end{subfigure}
\begin{subfigure}{.45\linewidth}
\includegraphics[width=\linewidth]{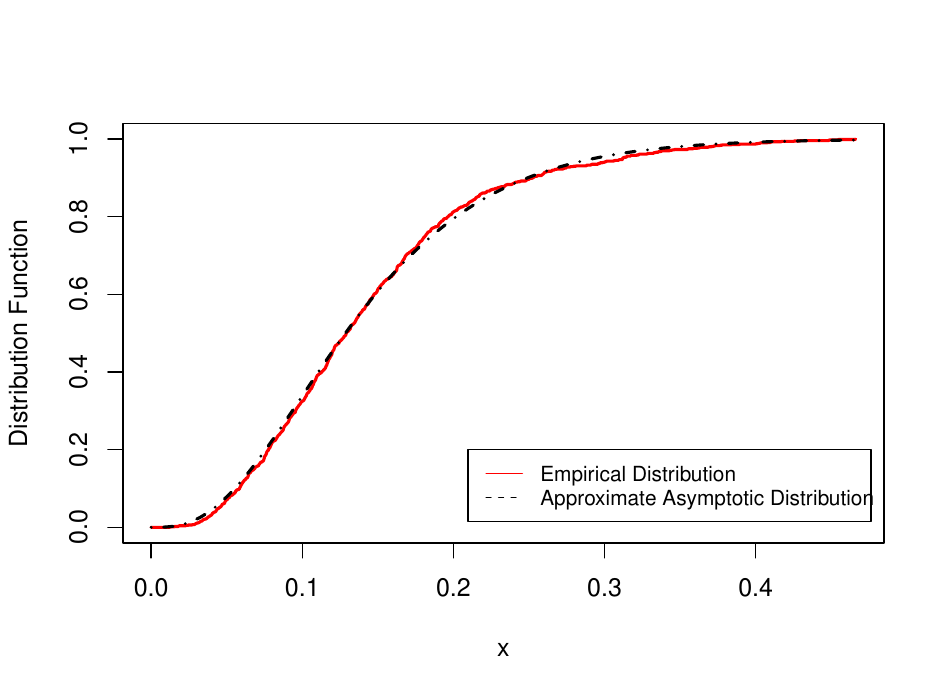}
\caption{$N=210$}
\end{subfigure}
\caption{Empirical Distribution of $\sum_{q=1}^3 D^2\left(F_{q},F;\linearK\right)$ (red solid curve) v.s. the approximate Gamma distribution (black dashed curve).}
\label{fig:AsympDistL3}
\end{figure}

\noindent{\bf Example 2}. Using the same $\mZ$ as in Example 1, we consider the case when $L=3$.
Figure \ref{fig:AsympDistL3} compares the empirical distribution of $\sum_{q=1}^3  D^2\left(F_q,F;\linearK\right)$ of 1000 random and balanced assignments and the approximate distribution $\text{Gamma}\left(\frac{2d}{u},\frac{3u}{N}\right)$ for $N=30, 60, 120,$ and 210.

Extending the linear kernel to a higher order, we obtain the polynomial kernel, defined by 
\[
K_{\text{Poly}}(\vz,\vt)=(1+\vx^\top \vt)^r,
\]
where $\vx, \vt\in \Omega$ and $r$ is the polynomial degree, and thus $r=1$ for $\linearK$  and $r=2$ for quadratic kernel. 
The RKHS $\mathcal{H}$ includes all the polynomial basis functions of the covariates up to the same degree $r$. 
Although more general than the linear kernel, the corresponding discrepancy criterion does not have a trackable asymptotic distribution. 
To decide the threshold of the rerandomization, we need to first generate a large number of random assignments, compute the corresponding polynomial discrepancies, and obtain the empirical distribution. 
Based on it, we can select the sample percentile as the threshold. 

\subsection{Linear Discrepancy v.s. Mahalanobis Distance}\label{subsec:LinearvsMD}

\cite{morgan2012rerandomization} and \cite{morgan2015rerandomization} have proposed to use Mahalanobis Distance as the rerandomization criterion. 
Using the notation in this paper, MD for $L=2$ defined in \cite{morgan2012rerandomization} is 
\[
\text{MD}:= \frac{N}{4} (\bar{\bm Z}^{(1)}-\bar{\bm Z}^{(2)})^\top \cov(\mZ)^{-1}(\bar{\bm Z}^{(1)}-\bar{\bm Z}^{(2)}),
\]
where $\bar{\bm Z}^{(q)}$ is the vector of sample means of the $d$-dimensional covariates for the $q$-th treatment group, i.e., $(\bar{Z}_1^{(q)}, \ldots, \bar{Z}_d^{(q)})$ and $\cov(\mZ)$ is the sample covariance matrix of all the covariates. 
It is easy to see that the MD is the same as the $(N/4)D^2(F_1, F_2;\linearK)$ if all the covariates are standardized and \emph{decorrelated} such that $\cov(\mZ)^{-1}$ is an identity matrix.  
Also, \cite{morgan2012rerandomization} and \cite{morgan2015rerandomization} pointed out that since $(\bar{\bm Z}^{(1)}-\bar{\bm Z}^{(2)})$ is asymptotic normal and thus MD follows a Chi-square distribution asymptotically, which is also a special case of the Gamma distribution in Theorem \ref{thm:linearL3approx}. 

Given this connection with the existing MD as a rerandomization criterion, we need to emphasize the new contributions of this paper. 
First, the existing literature on MD-based rerandomization has focused on the $L=2$ case.
For $L\geq 3$, the recommendation has been vague. 
If directly extending the MD criterion to the general $L\geq 2$ case, we should compute all the pairwise MD criteria between every two treatment groups. 
Along this direction, we face two options to define the rerandomization criterion. 
One option is to require the sum of these $\binom{L}{2}$ MDs to be smaller than a certain threshold. 
The other option is to require every MD to be smaller than a certain threshold.
For the first option, it is necessary to derive an approximate distribution for the sum of the MDs, similarly to what we have done for the linear discrepancy previously. 
For the second option, procedures such as the Bonferroni method must be used to adjust $p$ (percentile in \eqref{eqn:rerandcri}) to determine the proper threshold for the simultaneous $\binom{L}{2}$ tests of MDs. 
All these issues have not yet been fully addressed in the literature.

The general rerandomization criterion proposed in this paper is for any $L\geq 2$, which fills this void in the literature. 
Theorem \ref{thm:discequality} in Section \ref{sec:rerandom} has shown that we only need to use the sum of $D^2(F_q,F; K)$ as the balancing criterion, which is equivalent to using the sum of all the pairwise $D^2(F_q, F_{q'}; K)$. 
More importantly, Theorem \ref{thm:discequality} holds for any positive definite kernel and not just for a linear kernel. 
Theorem \ref{thm:linearL3approx} provides an approximated distribution for the balancing criterion using the linear kernel, based on which the threshold $a_p$ can be decided from a known Gamma distribution. 
Second, although a minor point, calculating $D^2(F_1, F_2; \linearK)$ is simpler than the MD because there is no need to invert the sample covariance matrix $\cov(\mZ)$, which can be ill-conditioned. 
The parameters of the Gamma distribution are simple, and they only need to be computed once throughout the rerandomization procedure.

The linear kernel has some obvious advantages. 
The corresponding linear discrepancy is easy to compute. 
An approximate distribution of the criterion is available to decide the critical value for rerandomimzation. 
However, the linear kernel has limitations. 
First, it is not a universal kernel \citep{micchelli2006universal}, and the linear discrepancy can only distinguish the two distributions in terms of the mean values. 
Second, the corresponding RKHS $\mathcal{H}$ only consists of linear functions of all the covariates. 
Therefore, the basis function $f(\bm z)$ in \eqref{model:rm} in Assumption \ref{ass:model} can only be linear in $\bm z$, which is too restrictive and does not hold in many practical situations.

\subsection{$\mathcal{L}_2$-Discrepancy}\label{subsec:others}

As introduced in Section \ref{sec:framework}, the reproducing kernel $K$ should be chosen such that the corresponding RKHS $\mathcal{H}$ is large enough to include all possible basis functions $f(\bm z)$ in \eqref{model:rm}. 
It also improves the robustness of the treatment assignment to model uncertainty. 

Assume the domain of the covariates is bounded rectangle, $\Omega = [a_1,b_1]\times[a_2,b_2]\cdots\times[a_d,b_d]$, where $a_1,\cdots,a_d$ and $b_1,\cdots,b_d$ are the lower and upper bounds of the covariates $z_1,\cdots,z_d$, respectively. 
We consider a commonly used reproducing kernel defined on the rectangular domain, given by
\begin{equation}\label{eqn:kernelchoice}
K_{\mathcal{L}_2}(\vz,\bm t) = \prod\limits_{j=1}^d \left[1+\frac{1}{2}(|t_j|+|z_j|-|z_j-t_j|)\right].
\end{equation}
The corresponding RKHS $\mathcal{H}$ contains all functions whose mixed partial derivatives up to the first order are square integrable. 
The concrete formulas of the inner product and norms of this $\mathcal{H}$ can be found in \cite{li2020transformed}. 
This kernel function leads to the well-known $\mathcal{L}_2$-discrepancy \citep{hickernell1998generalized}, denoted by $D(\hat{F}, F_{\text{Uniform}};K_{\mathcal{L}_2})$, which is used to generate low-discrepancy sequence \citep{dick2010digital} in the area Quasim Monte Carlo.
It measures the difference between the empirical distribution function $\hat{F}$ of any samples from $\Omega=[0,1]^d$ and $F_{\text{Uniform}}$, the extract CDF of the uniform distribution in $\Omega$. 
Its closed form is available in \cite{hickernell1998generalized} and \cite{li2020transformed}. 

If using this kernel to define the covariate balancing criterion, the squared discrepancy $D^2(F_q, F;K_{\mathcal{L}_2})$ measures the difference between the empirical distribution functions $F_q$ and $F$. 
Note that $F$, the empirical distribution function of the covariates of all test units, is not the same as $F_{\text{Uniform}}$, so $D^2(F_q, F;K_{\mathcal{L}_2})$ is not the same as the original $\mathcal{L}_2$-discrepancy from \cite{hickernell1998generalized} and \cite{li2020transformed}. 
To compute $D^2(F_q, F;K_{\mathcal{L}_2})$, we need to follow \eqref{eq:discDefdiscrete}. 
However, we still name the rerandomization criterion, $\sum_{q=1}^L D^2(F_q, F;K_{\mathcal{L}_2})$, the $\mathcal{L}_2$-discrepancy because it is derived from the same kernel function. 

The advantage of the $\mathcal{L}_2$-discrepancy is that the corresponding RKHS $\mathcal{H}$ contains more complicated functions of the covariates than the linear or polynomial discrepancy, and polynomial functions are also in $\mathcal{H}$. 
Unfortunately, the distribution of $\sum_{q=1}^L D^2(F_q, F;K_{\mathcal{L}_2})$ with respect to the random assignment is not tractable.
Therefore, to obtain the threshold $a_p$, we can obtain the empirical distribution of $\sum_{q=1}^L D^2(F_q, F;K_{\mathcal{L}_2})$ from simulations. 
For example, we can generate $B=1000$ random assignments and compute the corresponding discrepancies, and then set $a_p$ as the $p\times 100\%$ sample percentile of the criterion values. 
As $N$ becomes larger, we should increase $B$.

To compare all the mentioned discrepancies, we construct the following example. 

\noindent{\bf Example 3.} We consider $N=20$ test units, each with $d=3$ covariates. The covariates for $n=10$ test units are generated as i.i.d. samples from $\mathcal{N}(-3\times {\bf 1},{\mI}_3)$, while the remaining 10  test units are drawn from $\mathcal{N}(5\times {\bf 1},{\mI}_3)$. 
There are $\binom{20}{10} = 184,756$ possible partitions, which can be handled by a standard laptop.
In this example, we compare the performance of rerandomization based on MD, as well as linear-, quadratic-, and $\mathcal{L}_2$-discrepancies. 
For each partition, we calculate the values of all four rerandimization criteria.

For all four types of rerandomization criterion, we use the true $0.05\times 100\%$ percentile of all possible partitions as the thresholds. 
This results in approximately 9,238 (5\%) qualified rerandomized assignments for each discrepancy. 
The $a_{0.05}$ percentile values for MD, linear-, quadratic-, and $\mathcal{L}_2$-discrepancies are 0.6134, 0.4293, 2.9331, and 2.5755, respectively. 
 
The results are presented in Table \ref{tab:compare_upper}. 
Each row is for one type of treatment assignment approach. 
For example, in the last row, we check all the random assignments that are acceptable by the $\mathcal{L}_2$-discrepancy criterion, and compute their means of the $D^2\left(F_1,F_2;K\right)$ using three kernels.
The first row is the average squared discrepancies using three kernels, but for all randomizations. 
Note that $\E\left\{\left.D^2\left(F_1,F_2;K\right)\right|\mZ,\vX \right\}$ is the part of (not the entire) the upper bound in \eqref{ineqn:discrepancy} of the variance of the difference-in-means estimator. 

\begin{table}[htb]
\centering
\caption{Mean-Squared Discrepancy of randomized assignments using different criteria under different kernels.}
\begin{tabular}{lccc}
\hline
&        \multicolumn{3}{c}{$\E\left\{\left.D^2\left(F_1,F_2;K\right)\right|\mZ,\vX \right\}$}  \\
\cline{2-4}
& \multicolumn{1}{c}{$K_{\text{Linear}}$} & \multicolumn{1}{c}{$K_{\text{Quadratic}}$} & \multicolumn{1}{c}{$K_{\mathcal{L}_2}$} \\
\hline
Completely randomization & 317.20 & 5641.78 & 521.63 \\
MD rerandomization   & 0.24  & 19.87 & 6.72 \\
Linear-discrepancy rerandomization  & 0.08  & 21.37 & 6.30 \\
Quadratic-discrepancy rerandomization  & 0.71  & 4.12  & 6.50 \\
$\mathcal{L}_2$-discrepancy rerandomization   & 0.21  & 9.30  & 5.08 \\
\hline
\end{tabular}%
\label{tab:compare_upper}%
\end{table}%

It is expected that a certain type of rerandomization achieves the smallest mean-squared discrepancy under the same kernel discrepancy that defines the criterion. 
For example, the linear-discrepancy rerandomization has the smallest $\E\left\{\left.D^2\left(F_1,F_2;K\right)\right|\mZ,\vX \right\}$ when $K$ is $K_{\text{Linear}}$. 
However, the $\mathcal{L}_2$-discrepancy serves as a more conservative and robust criterion, attaining the second smallest $\E\left\{\left.D^2\left(F_1,F_2;K\right)\right|\mZ,\vX \right\}$ for linear and quadratic kernels, and of course it has the smallest mean-squared discrepancy when $K$ is $K_{\mathcal{L}_2}$.  
Moreover, it ensures a small worst-case estimate variance when the relationship between covariates and response is more complex.
We recommend that when the linear or quadratic assumptions for $f(\vz)$ are not certain, the $\mathcal{L}_2$-discrepancy is a more robust rerandomization criterion. 

\section{Rerandomization for Factorial Experiment}

In this section, we focus on the controlled experiments in which the $L$ treatment settings are constructed by a fractional or full factorial design. 
Theorem \ref{thm:discequality} and Corollary \ref{cor:discequality} provide a smart strategy to generate discrepancy-based rerandommization of $N$ test units into $L$ treatment groups. 
We first focus on the two-level factorial design, as it is among the most basic and yet useful experimental design methods \citep{wu2011experiments}. 
Briefly, our idea is to partition the total $N$ test units into $L= 2^s$ groups following an $s$-depth binary tree.

\begin{figure}[htb]
\centering
\includegraphics[width=0.8\textwidth]{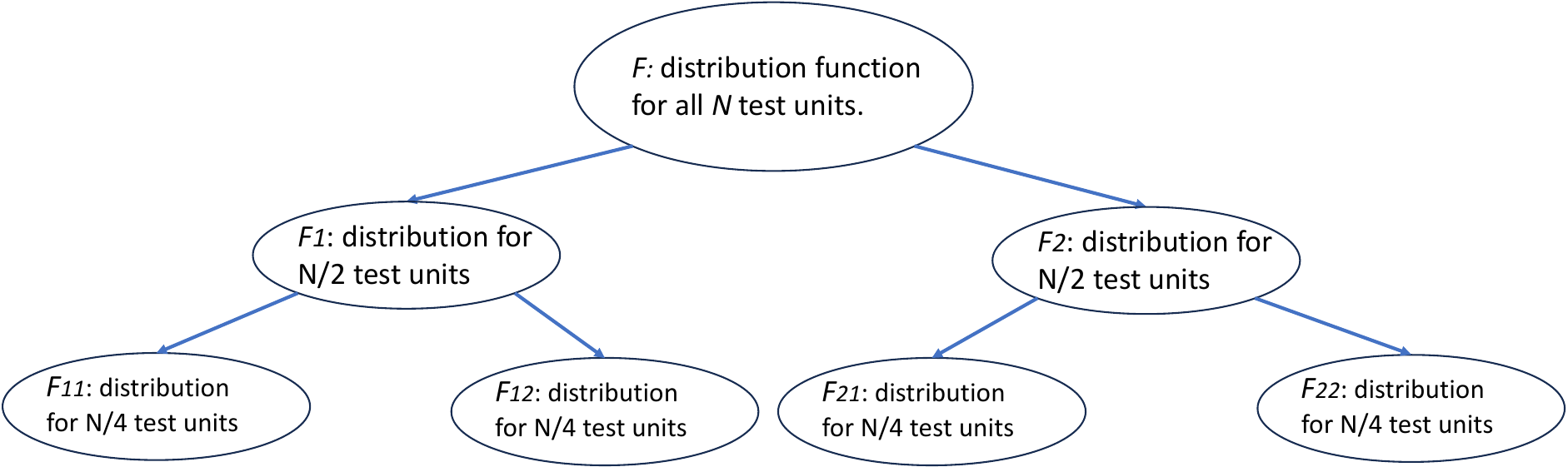}
\caption{Partition $N$ test units into $L=2^s$ ($s=2$) groups via $s$-depth binary tree.\label{fig:tree}}
\end{figure}

We illustrate the proposed strategy using the simple $L=2^2=4$ case. 
Figure \ref{fig:tree} shows a $2$-depth binary tree. 
The root of the tree represents all $N$ test units with covariates following the empirical distribution function $F$. 
From any random assignment for $L=2$, we can partition the $N$ test units into the two groups, represented by the two nodes in the first layer in Figure \ref{fig:tree}. 
In each of the two nodes, there are $N/2$ test units whose covariates follow the empirical distribution function $F_1$ or $F_2$. 
Then we proceed to use two different random assignments for $L=2$ and $N/2$ to split the two nodes. 
As shown in Figure \ref{fig:tree}, we obtain four end nodes. 
Each has $N/4$ test units whose covariates follow one of the four empirical distribution functions $F_{11}$, $F_{12}$, $F_{21}$, and $F_{22}$. 

The final assignment $\vX$ is composed of three two-level random assignments for $N$ and $N/2$ test units. 
Why should we consider this approach since it is certainly more direct to use a single discrepancy rerandomization for $L=4$?
As explained in \cite{morgan2015rerandomization}, it takes more time to simulate acceptable assignments for larger $N$ and small $a_p$. 
If we need to compute a more complicated discrepancy criterion, this process can take longer. 
For the same $N$, it is easier to generate acceptable rerandomized assignments for $L=2$ than for $L=4$, given the same value of $p$ in $a_p$.  
For instance, if $N=12$, it is faster to simulate $5\%$ accepted assignments from $\binom{12}{6}=924$ possible assignments when $L=2$ than from the $(12)!/(3!3!3!3!)=369,600$ possible assignments when $L=4$.

The proposed composite assignment following the binary tree can achieve a small discrepancy criterion, as long as each of the three two-level assignments meets the rerandomization criterion. 
We define $\phi(\mZ,\bm X)=1$, if the composite assignment $\vX$ satisfy the following
\begin{align*}
D^2(F_1, F_2; K) \leq a^{(1)}, \quad D^2(F_{11}, F_{12}; K)\leq a^{(21)}, \quad D^2(F_{21}, F_{22}; K)\leq a^{(22)}. 
\end{align*}
The thresholds $a^{(1)}$, $a^{(21)}$, and $a^{(22)}$ are to be discussed later. 

Based on the definition of the discrepancy criterion in \eqref{eqn:rerandcri}, 
\[\sum_{i=1}\sum_{j=1} D^2(F_{ij}, F; K)=\frac{1}{4}\left[\sum_{i=1}^2 D^2(F_{i1}, F_{i2};K)+\sum_{j,j'=1,j\neq j'} D^2(F_{1j}, F_{2j'};K)\right].\]
We only need to check the discrepancy $D(F_{1j}, F_{2j'}; K)$ for any two end nodes that are from different parent nodes. 
Recall in Section \ref{sec:preliminary}, we have reviewed that kernel discrepancy can also be interpreted as the norm of measures defined in \eqref{eq:discDefmeasure}. 
So it should satisfy the triangle inequality as all norms do. 
Let $\nu_{*}$ be the measure corresponding to the empirical distribution function $F_{*}$ of the same subscript `$*$'. 
In our case, these measures are the normalized sum of Dirac measures. 
Therefore, for $j,j'=1,2$ and $j\neq j'$,
\begin{align*}
D(F_{1j}, F_{2j'}; K)&=\norm[\cm]{\nu_{1j} - \nu_{2j'}}= \norm[\cm]{\nu_{1j} -\nu_1+\nu_1-\nu_2+\nu_2-\nu_{2j'}}\\
&\leq \norm[\cm]{\nu_{1j} - \nu_{1}}+\norm[\cm]{\nu_{1} - \nu_{2}}+\norm[\cm]{\nu_{2j'} - \nu_{2}}\\
&=D(F_{1j}, F_{1}; K)+D(F_{1}, F_{2}; K)+D(F_{2j'}, F_{2}; K).
\end{align*}
Based on Corollary \ref{cor:discequality}, we know $D(F_{1j}, F_{1}; K)=\frac{1}{2}D(F_{11}, F_{12}; K)\leq \frac{1}{2}\sqrt{a^{(21)}}$, $D(F_1, F_2; K)\leq \sqrt{a^{(1)}}$, and $D(F_{2j'}, F_{2}; K)=\frac{1}{2}D(F_{21}, F_{22}; K)\leq \frac{1}{2}\sqrt{a^{(22)}}$
and thus 
\[D(F_{1j}, F_{2j'}; K) \leq \sqrt{a^{(1)}}+\frac{1}{2}\left(\sqrt{a^{(21)}}+\sqrt{a^{(22)}}\right).\]
The discrepancy rerandomization criterion in \eqref{eqn:rerandcri} is upper bounded by 
\begin{align}\nonumber
& \sum_{i=1}\sum_{j=1} D^2(F_{ij}, F; K)=\frac{1}{4}\left[\sum_{i=1}^2 D^2(F_{i1}, F_{i2};K)+\sum_{j,j'=1,j\neq j'} D^2(F_{1j}, F_{2j'};K)\right]\\
&\leq \frac{1}{4}\left[a^{(21)}+a^{(22)}+4\left(\sqrt{a^{(1)}}+\frac{1}{2}(\sqrt{a^{(21)}}+\sqrt{a^{(22)}})\right)^2\right]. \label{eq:2factorUB}
\end{align}
Therefore, if $a^{(1)}$, $a^{(21)}$, and $a^{(22)}$ are small enough, the overall threshold for the entire discrepancy is also small, ensuring the covariate balancing. 

This strategy can be easily extended to any $L=2^s$,  $L=3^s$, or even more general factorial design. 
For example, for $L=2\times 3$, we first use a discrepancy rerandomized assignment for $L=2$ and partition $N$ test units into two groups of size $N/2$, and then use two discrepancy rerandomized assignments for $L=3$ and partition each group into three sub-groups of size $N/6$. 
The final six groups should achieve a small discrepancy criterion when all the thresholds are sufficiently small as well. 

\noindent{\bf{Example 4}.} This example demonstrates the rerandomization strategy for the factorial experiment described above, using a dataset from the New York Department of Education (NYDE) in \cite{branson2016improving}.
The dataset contains information on 50 covariates for $N=1,376$ schools in 2008. Following \cite{branson2016improving}, nine possibly correlated covariates are considered: total enrollment, proportions of students in five racial categories (White, Black, Asian, Native American, and Latino), proportion of female students, enrollment rate, and poverty rate. 
This experiment aims to evaluate the effects of five ``incentive programs'' intended for high schools ``that desperately need performance improvement.'' 
A $L=2^5$ factorial design is used for the five factors, with $n=N/L=43$ schools assigned to each of the 32 treatment settings, resulting in $\binom{32}{2} = 496$ pairwise comparisons. 
A linear relationship between the covariate and response is reasonably assumed. Accordingly, we use the linear discrepancy for rerandomization. 
The $5\%$ percentile is used for all thresholds for the composite rerandomization procedure. 

\begin{figure}[tb]
\centering
\includegraphics[width=0.5\textwidth]{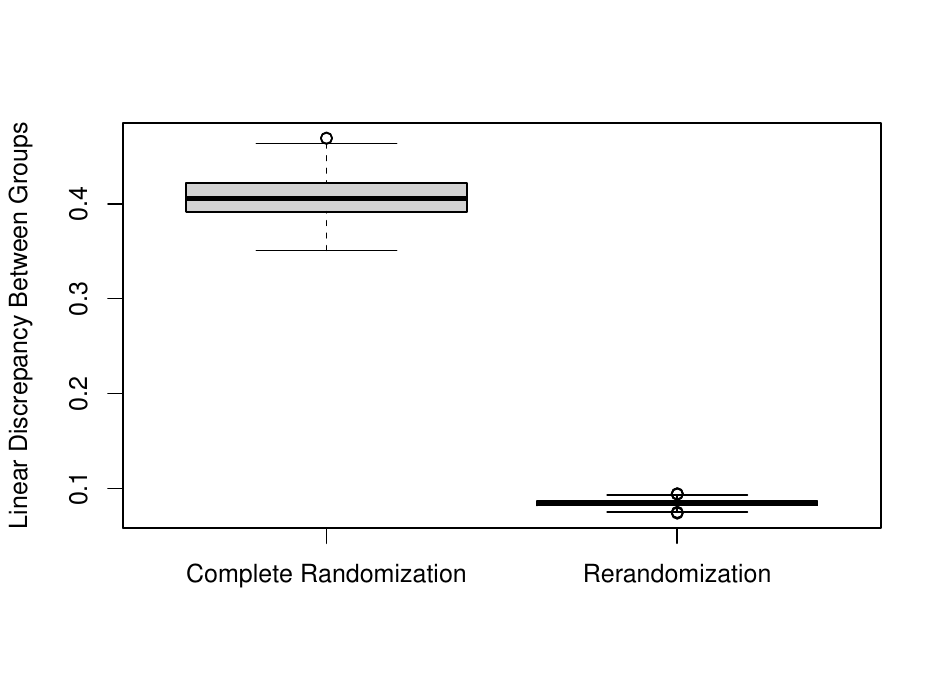}
\caption{Boxplot of $\E\left[\left.D^2(F_{i},F_{j};K)\right|\mZ,\phi(\mZ,\vX)=1\right]$ for the 496 pairs of treatment effects.}
\label{fig:NYDEBoxplot}
\end{figure}

We generate $B=100$ completely randomized assignments and $B=100$ rerandomized assignments following the proposed strategy. 
For each pair of $\alpha_i-\alpha_j$, we compute the average of the $B=100$ values of $D^2(F_i,F_j;\linearK)$, i.e., an estimate of $\E\left[D^2(F_i,F_j;\linearK)|\mZ,\phi(\mZ,\vX)=1\right]$, for the completely randomized assignments and rerandomized assignments.
It is part of the upper bound of the variance of $\hat{\alpha}_i-\hat{\alpha}_j$. 
We do so for all 496 pairs of treatment effects. 
Figure \ref{fig:NYDEBoxplot} displays the boxplot of the 496 mean $D^2(F_i,F_j;\linearK)$ values. 
The discrepancies of the proposed rerandomization method are substantially smaller and less variable than those of complete randomization.

\section{Numerical Examples}\label{sec:examples}
In this section, we use simulated data to compare complete randomization and rerandomization based on MD, linear, quadratic, and $\mathcal{L}_2$-discrepancies. 

\noindent {\bf Example 5}. We assume $f(\bm z)$ in \eqref{model:rm} is a linear function with $d=5$ covariates defined by
\begin{equation*}
f(\bm z_i) = \sum_{j=1}^d\beta_jz_{ij} ,\quad \text{ for }i=1,\ldots,N.
\end{equation*}  
The covariate matrix $\mZ = (\bm z_1,...,\bm z_N)^\top$ is generated as follows.
For the first two columns $\bm Z_1$ and $\bm Z_2$, we generate $70\%$ of the i.i.d. samples from $\mathcal{N}(\bm\mu_1,\Sigma)$, and $30\%$ from $\mathcal{N}(\bm\mu_2,\Sigma)$, where $\bm\mu_1=-3\times{\bf 1}$,  $\bm\mu_2=5\times {\bf 1}$, and the matrix $\Sigma$ is a random positive definite matrix.
For $\bm Z_3$, we generate $40\%$ of the i.i.d. samples from Uniform$(-0.5,1.5)$ and $60\%$ from Uniform$(-3,8)$.
For $\bm Z_4$, we generate $20\%$ of the i.i.d. samples from Gamma$(0.1,1)$ and $80\%$ from Gamma$(2.5,1)$.
For $\bm Z_5$, we generate $30\%$ of the samples from $\mathcal{N}(0.05,1)$ and $70\%$ from $\mathcal{N}(10,1)$.
The regression coefficients $\beta_1,\cdots,\beta_5$ are independently drawn from $\mathcal{N}(2,0.1)$, with each coefficient randomly assigned a positive or negative sign.

\begin{figure}[htb]
\centering
\begin{subfigure}{.45\linewidth}
\includegraphics[width=\linewidth]{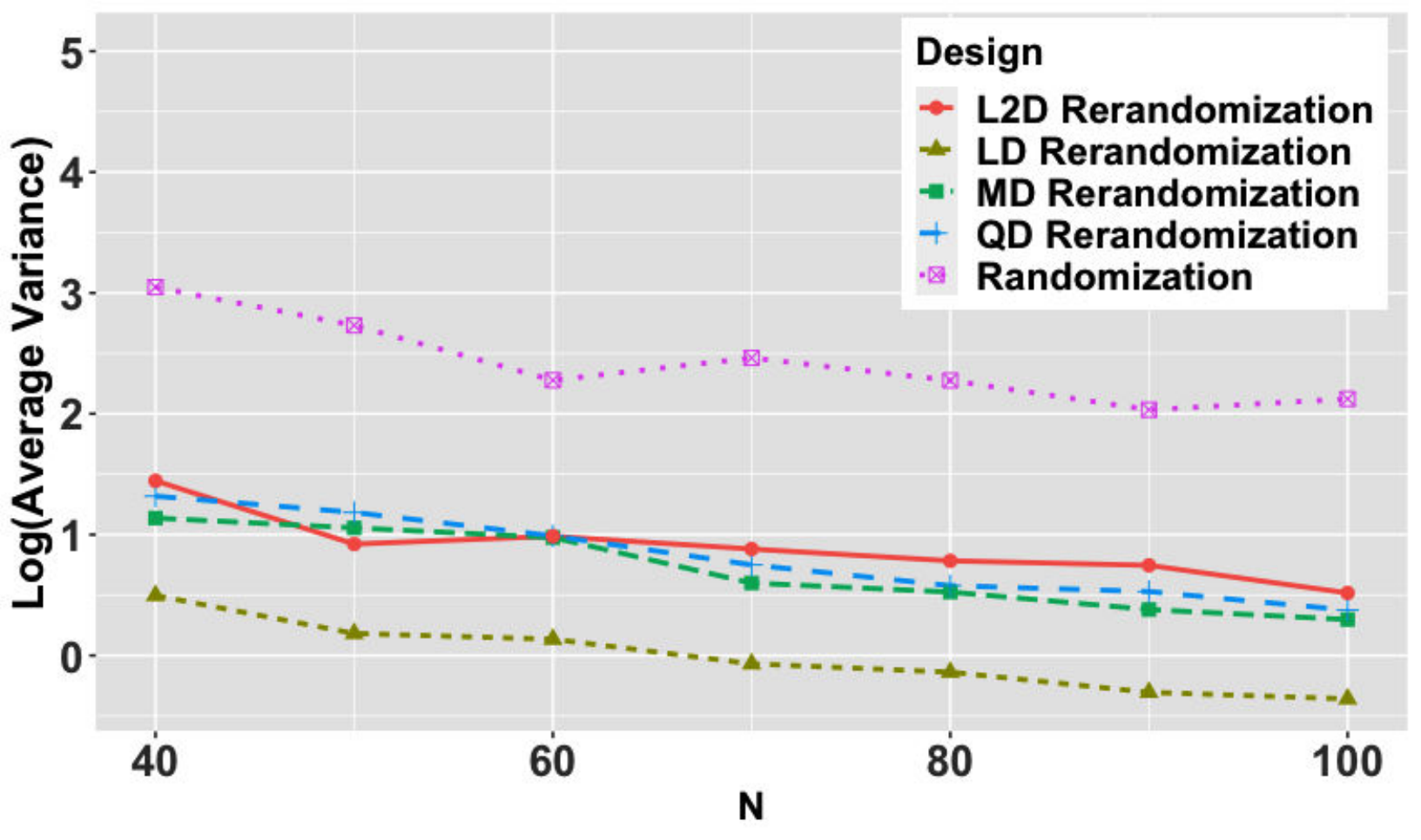}
\caption{$L=2$, Averaged Variance}
\end{subfigure}
\begin{subfigure}{.45\linewidth}
\includegraphics[width=\linewidth]{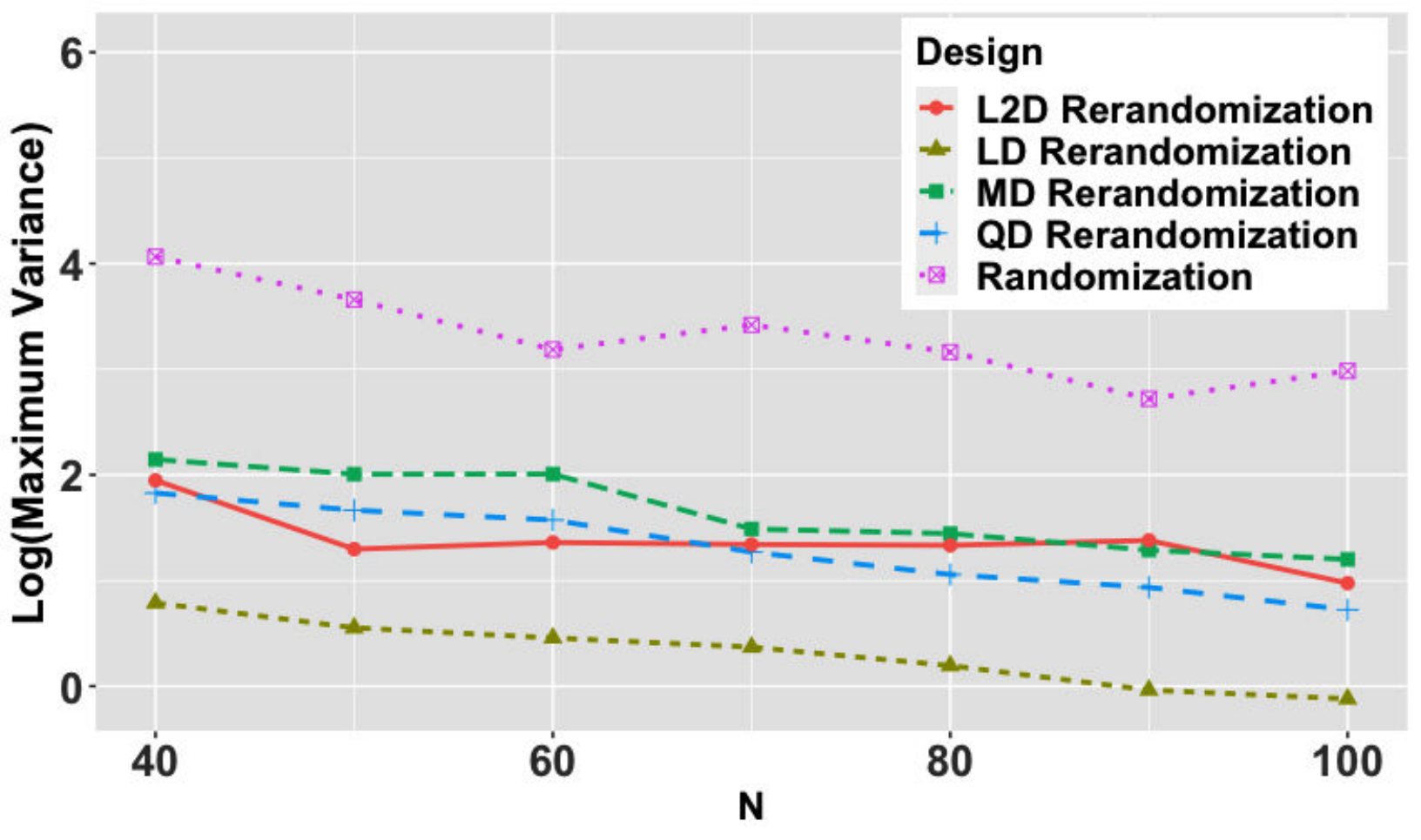}
\caption{$L=2$, Worst Variance}
\end{subfigure}
\begin{subfigure}{.45\linewidth}
\includegraphics[width=\linewidth]{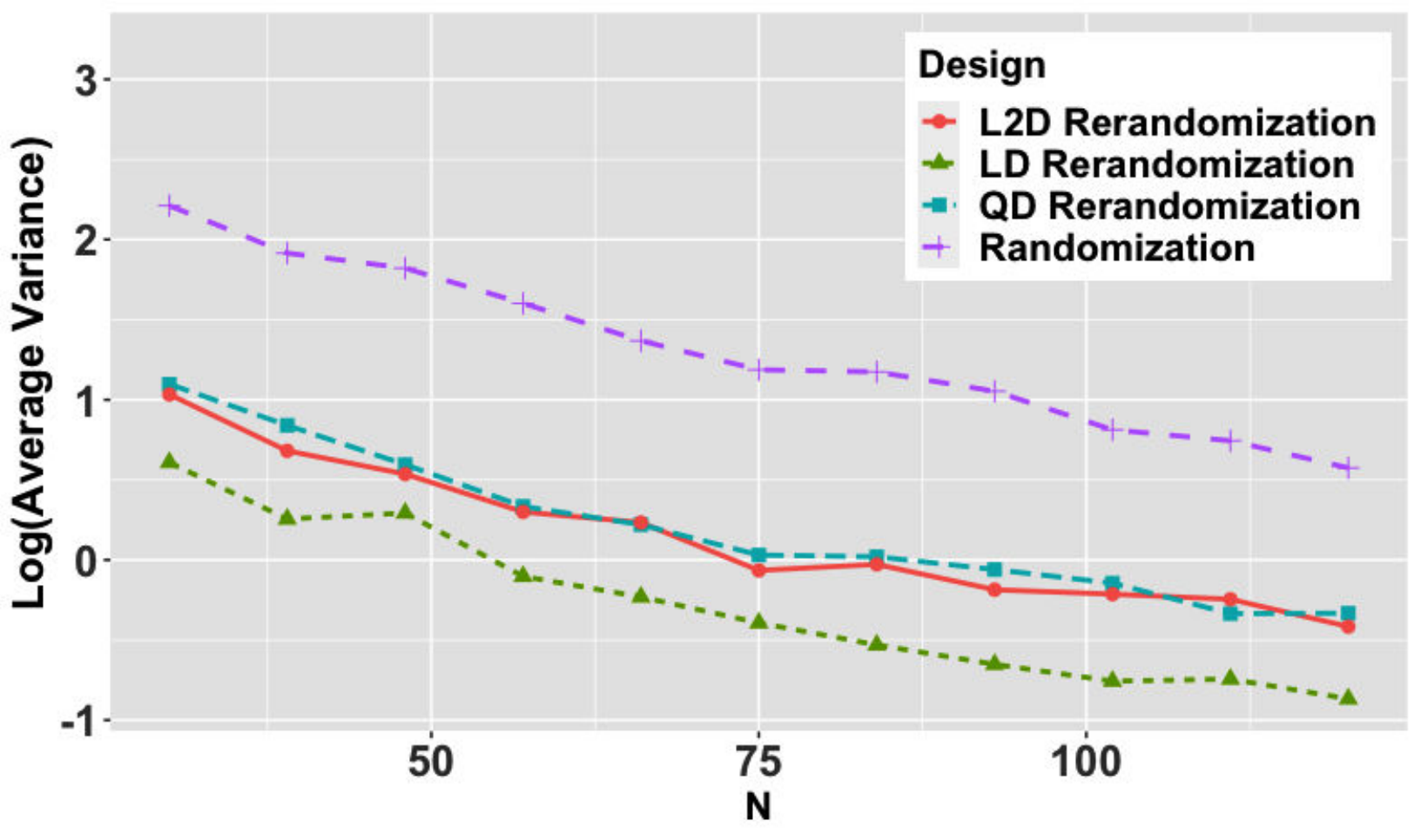}
\caption{$L=3$, Averaged Variance}
\end{subfigure}
\begin{subfigure}{.45\linewidth}
\includegraphics[width=\linewidth]{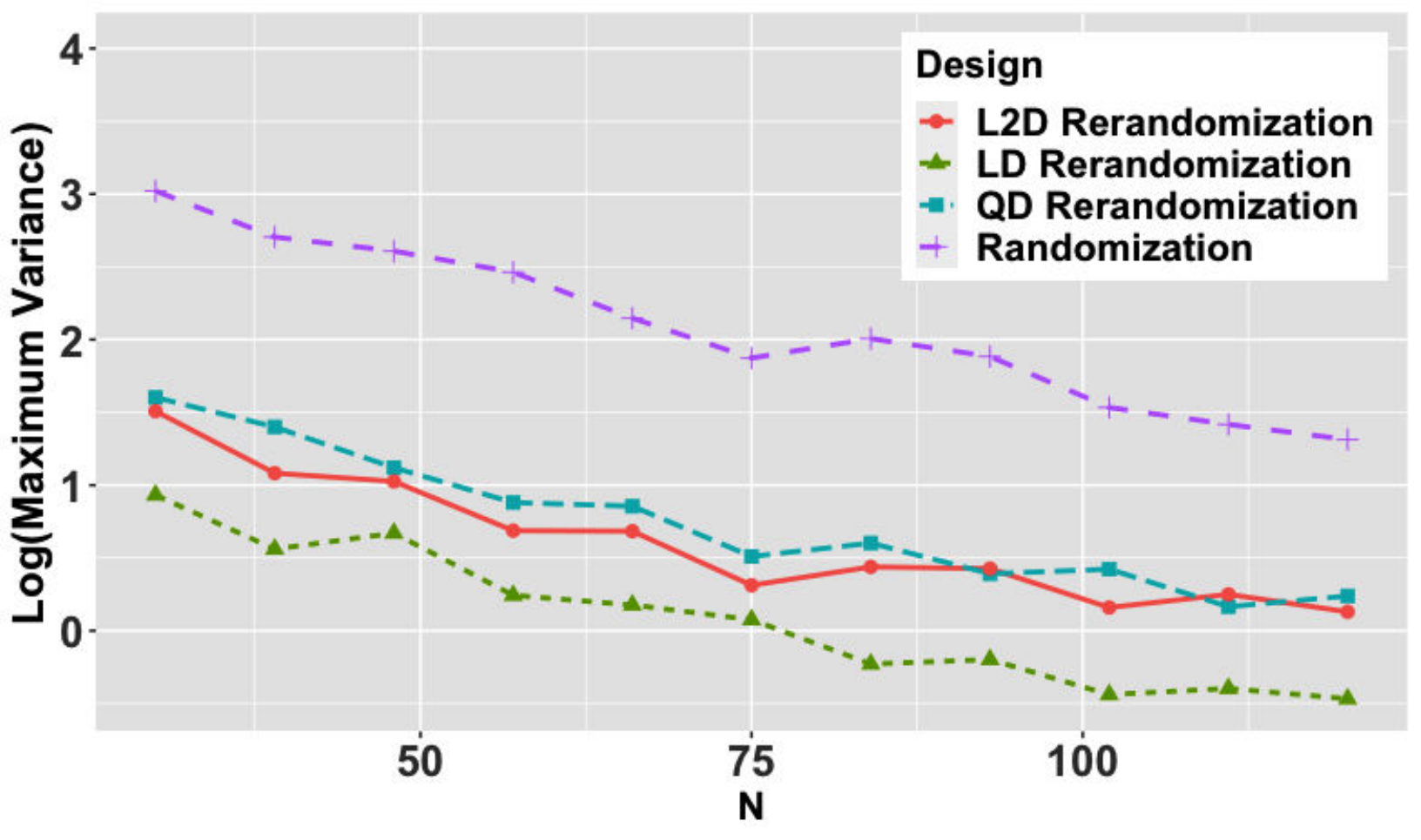}
\caption{$L=3$, Worst Variance}
\end{subfigure}
\caption{Average and the worst-case estimated variance of the difference-in-means estimator using different random assignments for $L=2$ and $L=3$.}
\label{fig:5dvarlinear}
\end{figure}

To compare the variance of the difference-in-means estimator under different strategies, $B=100$ assignments for each type are generated, and the estimated variance is calculated using \eqref{eqn:variance}. 
We compare the average of the estimated variance over $S=100$ randomly generated sets of regression coefficients. 
Figure \ref{fig:5dvarlinear} presents the averaged variance and the largest (or the \emph{worst-case}) variance of the difference-in-means estimator under the five different strategies with sample size $N$ varying from 40 to 100 for $L=2$, and from 30 to 120 for $L=3$. 
Since the asymptotic distribution of the summation of MD is unavailable for $L=3$, we omit it from the comparison. 

It is expected that the rerandomization based on linear discrepancy returns the smallest average and the smallest worst estimated variance since the underlying function $f$ is linear in the covariates. 
The complete randomization ranks at the bottom as it does not require any covariate balancing. 
The advantage of linear discrepancy over the $\mathcal{L}_2$-discrepancy and quadratic discrepancy is less significant for $L=3$, which might be due to the limited sample size. 
For the linear discrepancy, we set the threshold value $a_p$ from the approximated Gamma distribution in Theorem \ref{thm:discfiniteapprox}. 
The percentile $p$ is the same $5\%$ for all rerandomization strategies. 

{\bf Example 6}. In this example, we evaluate the performance of the proposed rerandomization when the input variables include both continuous and categorical variables with interactions, for experiments with higher levels. 
Let $f(\bm z)$ be a linear function of the main and two-factor interactions of $d=3$ covariates. 
\begin{equation*}
f(\bm z_i) =\beta_1z_{i1}+\beta_2z_{i2}+\beta_3z_{i3}+\beta_4z_{i1}z_{i2}+\beta_5z_{i1}z_{i3}+\beta_6z_{i2}z_{i3}\quad \text{ for }i=1,\ldots,N.
\end{equation*}
The columns $\bm Z_1$, $\bm Z_2$ are continuous covariates with 80\% of the entries in columns $\bm Z_1$ and $\bm Z_2$ are independently drawn from $\mathcal{N}(\bm\mu_1,\Sigma)$ and the remaining $20\%$ are drawn from $\mathcal{N}(\bm\mu_2,\Sigma)$, where $\bm\mu_1=-5\times{\bf 1}$ and  $\bm\mu_2=5\times {\bf 1}$. 
The covariance matrix $\Sigma$ is a random positive definite matrix. 
For the categorical variable $\bm Z_3$, half of its entries are randomly assigned 1, and the other half are set to 0. 
The regression coefficients $\beta_1,\cdots,\beta_6$ are independently drawn from $\mathcal{N}(2,0.1)$, with each coefficient randomly assigned a positive or negative sign.

\begin{figure}[htb]
\centering
\begin{subfigure}{.45\linewidth}
\includegraphics[width=\linewidth]{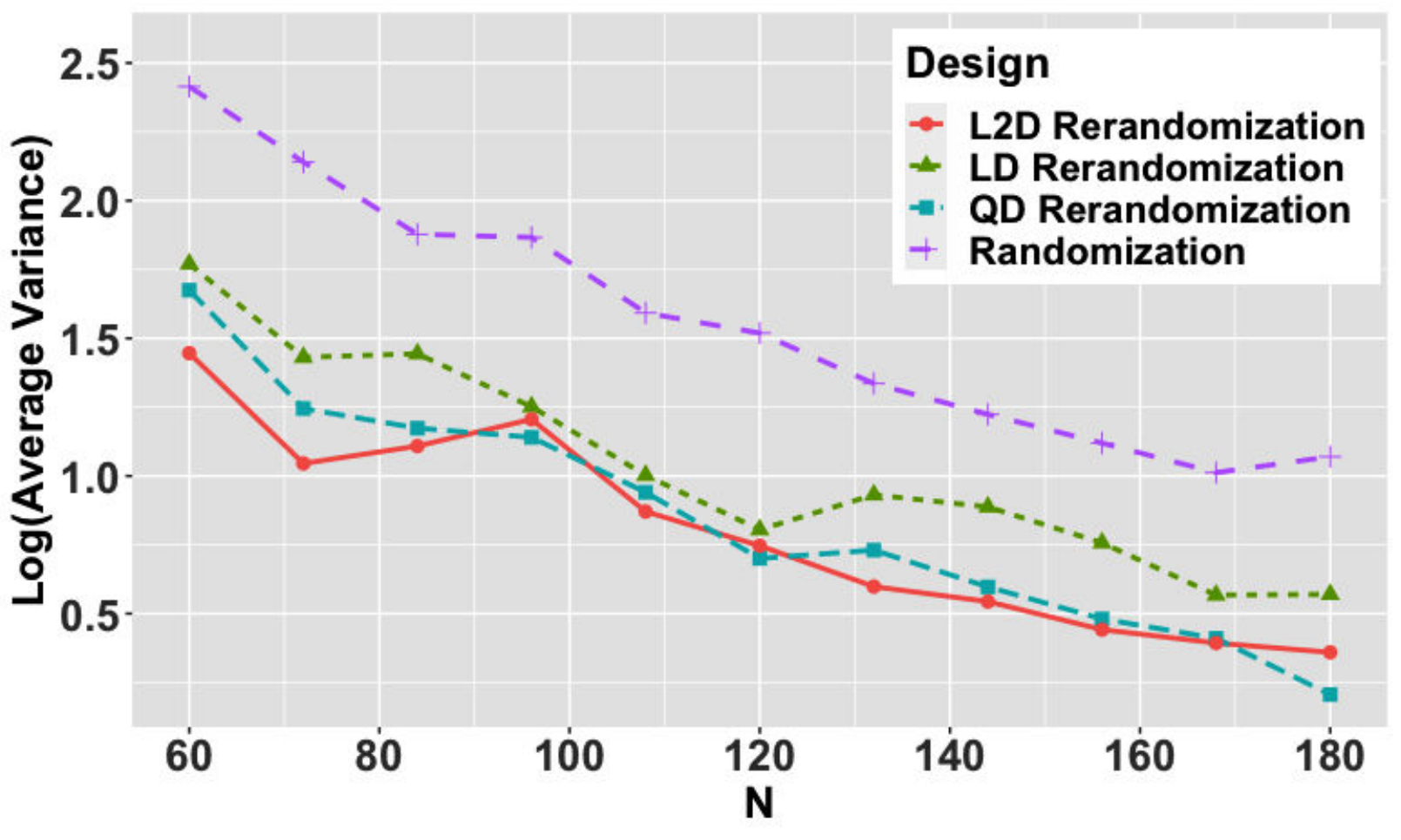}
\caption{$L=3$, Averaged Variance}
\end{subfigure}
\begin{subfigure}{.45\linewidth}
\includegraphics[width=\linewidth]{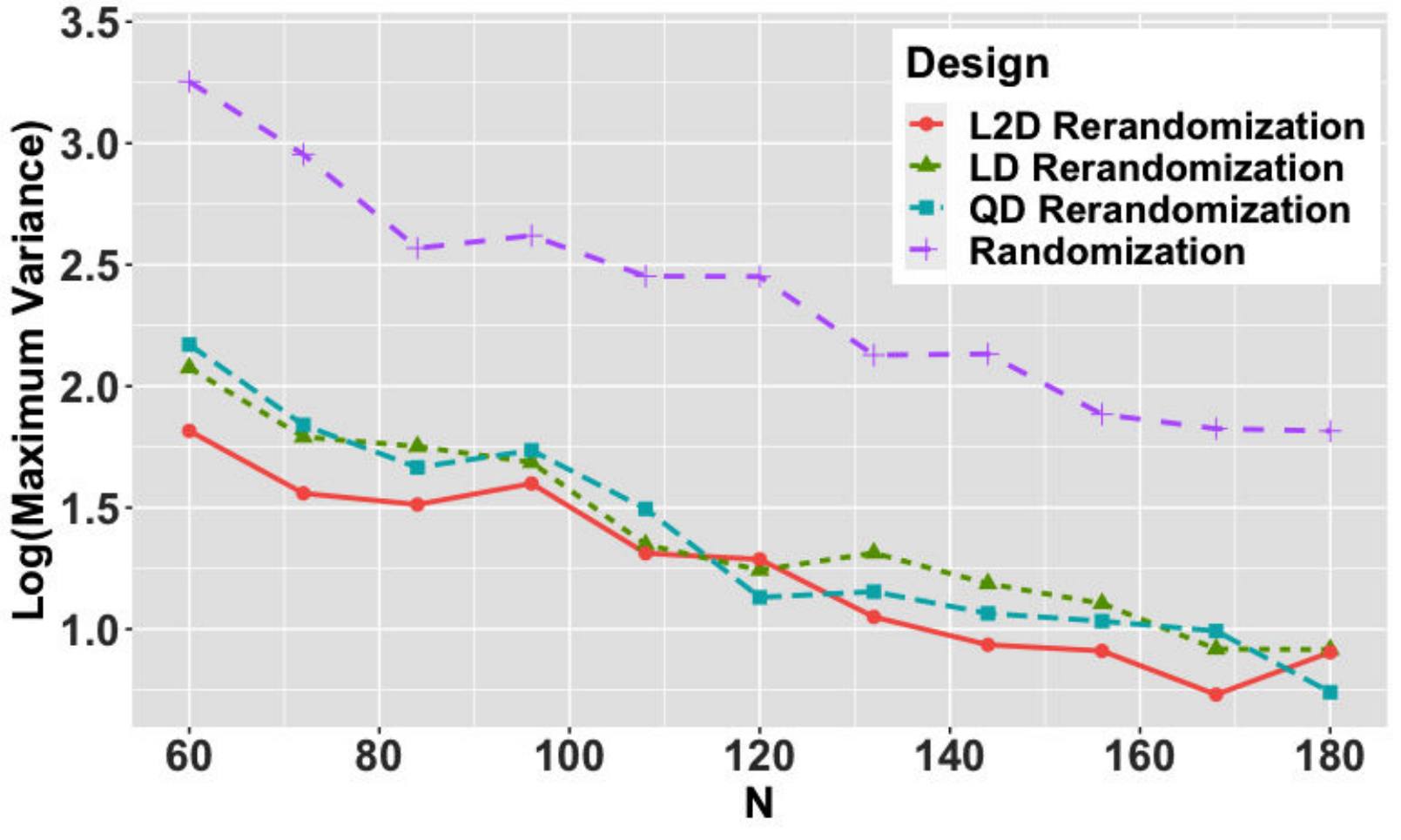}
\caption{$L=3$, Worst Variance}
\end{subfigure}
\begin{subfigure}{.45\linewidth}
\includegraphics[width=\linewidth]{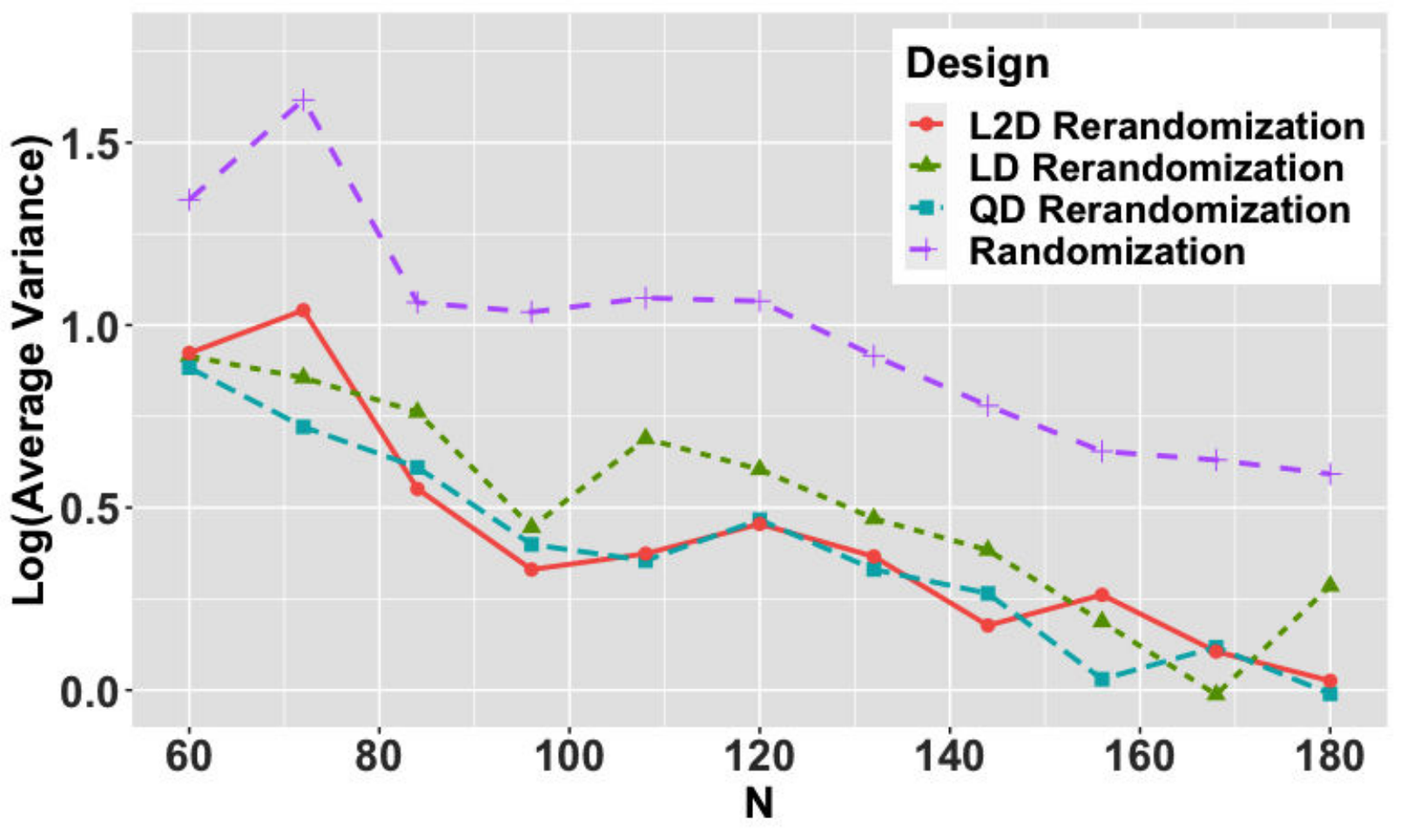}
\caption{$L=4$, Averaged Variance}
\end{subfigure}
\begin{subfigure}{.45\linewidth}
\includegraphics[width=\linewidth]{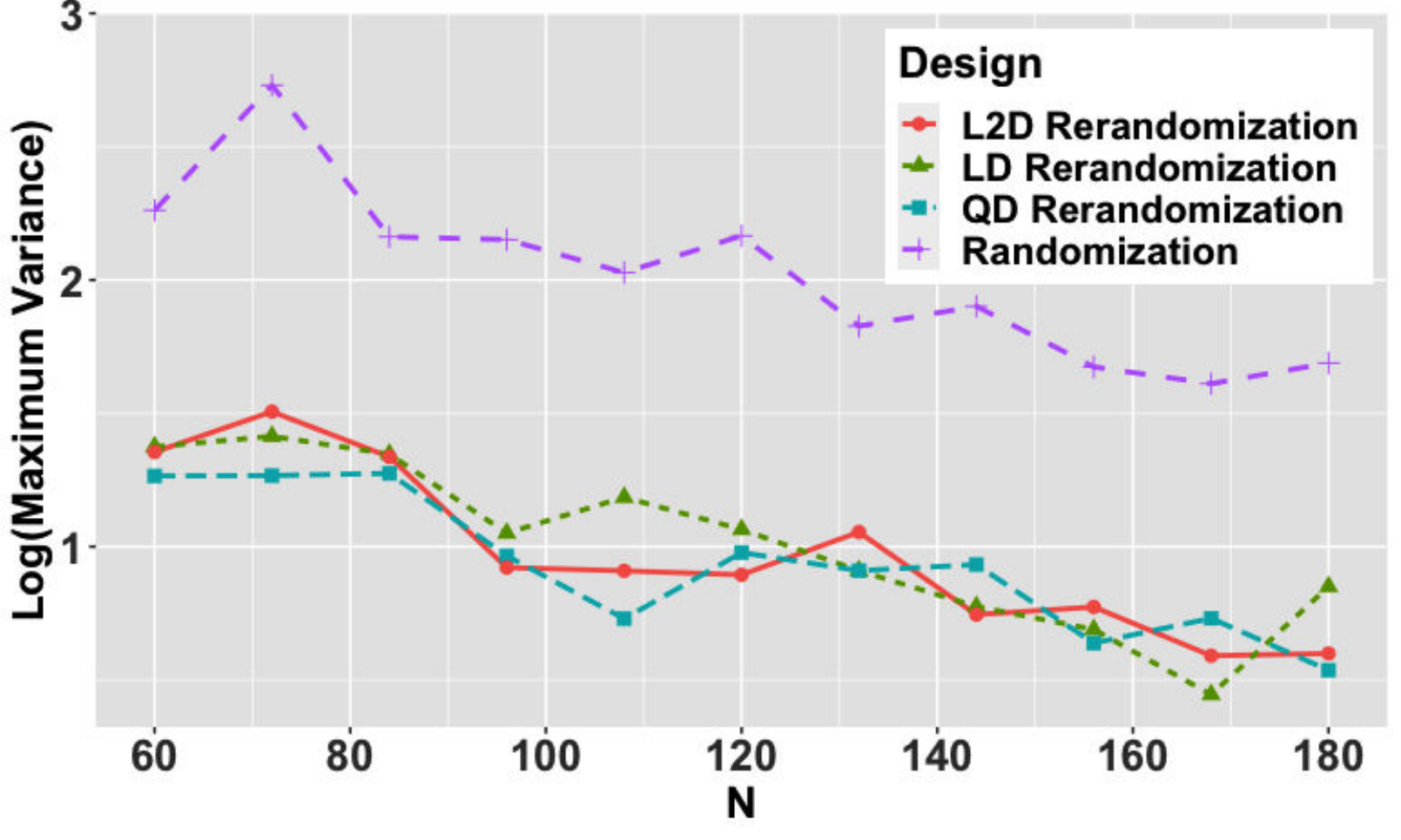}
\caption{$L=4$, Worst Variance}
\end{subfigure}
\caption{The average and worst-case estimated variance of the difference-in-means estimator using different random assignments with both continuous and binary covariates.}
\label{fig:3dvarlinearCate}
\end{figure}

As in Example 5, we compute and compare the average variance and worst-case variance across $B=100$ assignments for $S=100$ randomly generated sets of regression coefficients with sample size $N$ varying from 60 to 180, for $L=3$ and $4$. 
The threshold $a_p$ is the $0.05\times 100\%$ percentile of their respective distribution. 
For the linear discrepancy, we use the Gamma distribution. 
The comparison is presented in Figure \ref{fig:3dvarlinearCate}. 
MD is not included since $L=3$ and $4$. 
Among the four rerandomization strategies, $\mathcal{L}_2$ and quadratic discrepancy outperform the linear discrepancy, which is expected as $f(\bm z)$ has the 2nd order polynomial terms. 

{\bf Example 7}. In this example, we consider a non-polynomial function $f(\vz)$ \citep{lim2002design} of the $d=2$ covariate $\bm z\in [0,1]^2$, defined as
\[f(\bm z) = \frac{1}{6}[(30+5z_1\sin(5z_1))(4+\exp(-5z_2))-100],\,\,z_i\in[0,1], \text{for all }i=1,2.\]
Each row of the covariate matrix $\bm Z$ is independently sampled from $\text{Uniform}[0,1]$. 
We generate $B=500$ assignments using each of five treatment assignment strategies for $L=2$. 
For $L=3$, MD-rerandomization is omitted for the same reason as in Example 5. 
The sample size $N$ ranges from 100 to 200 for $L=2$ and 99 to 198 for $L=3$. 

\begin{figure}[htb]
\centering
\begin{subfigure}{.45\linewidth}
\includegraphics[width=\linewidth]{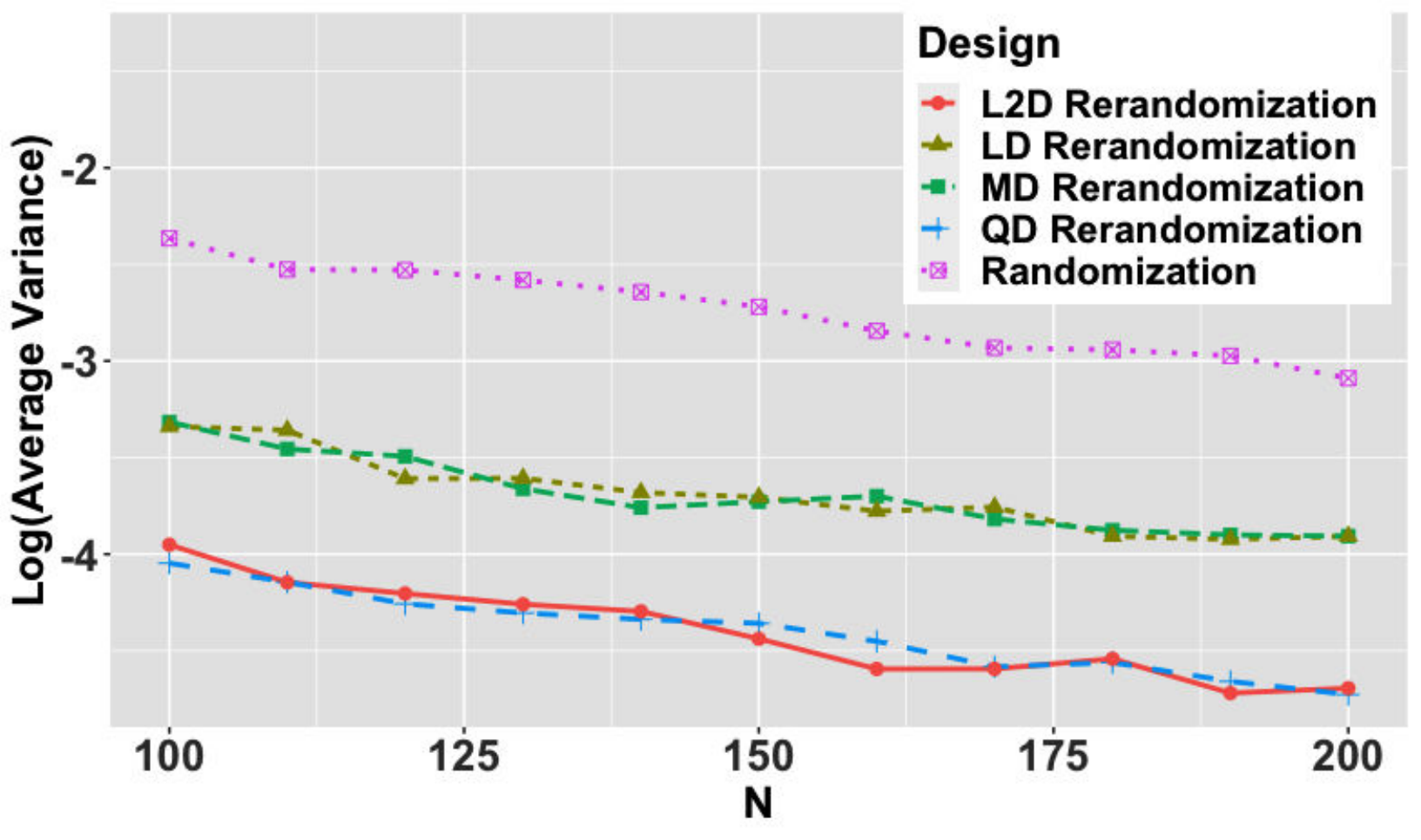}
\caption{$L=2$, Averaged Variance}
\end{subfigure}
\begin{subfigure}{.45\linewidth}
\includegraphics[width=\linewidth]{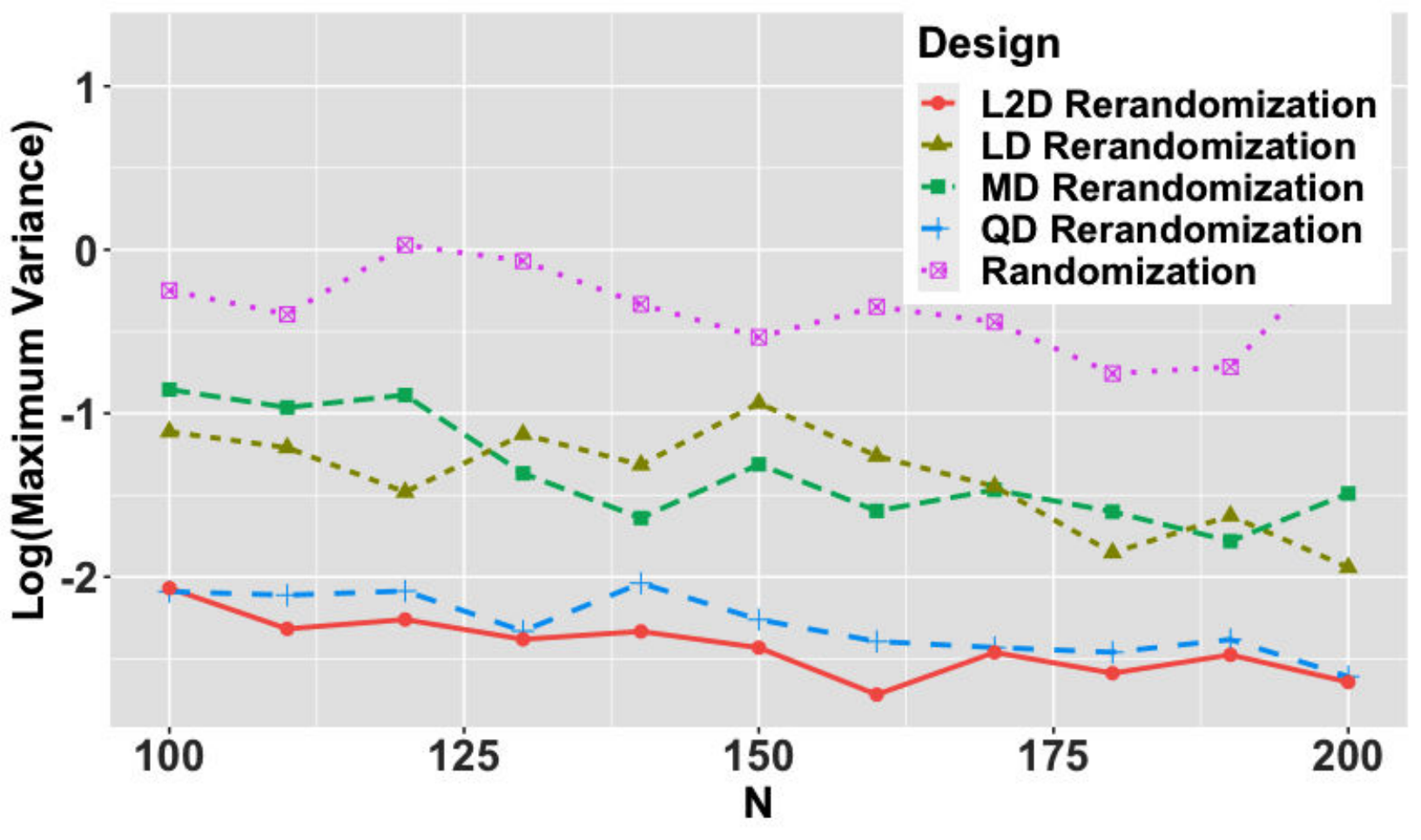}
\caption{$L=2$, Worst Variance}
\end{subfigure}
\begin{subfigure}{.45\linewidth}
\includegraphics[width=\linewidth]{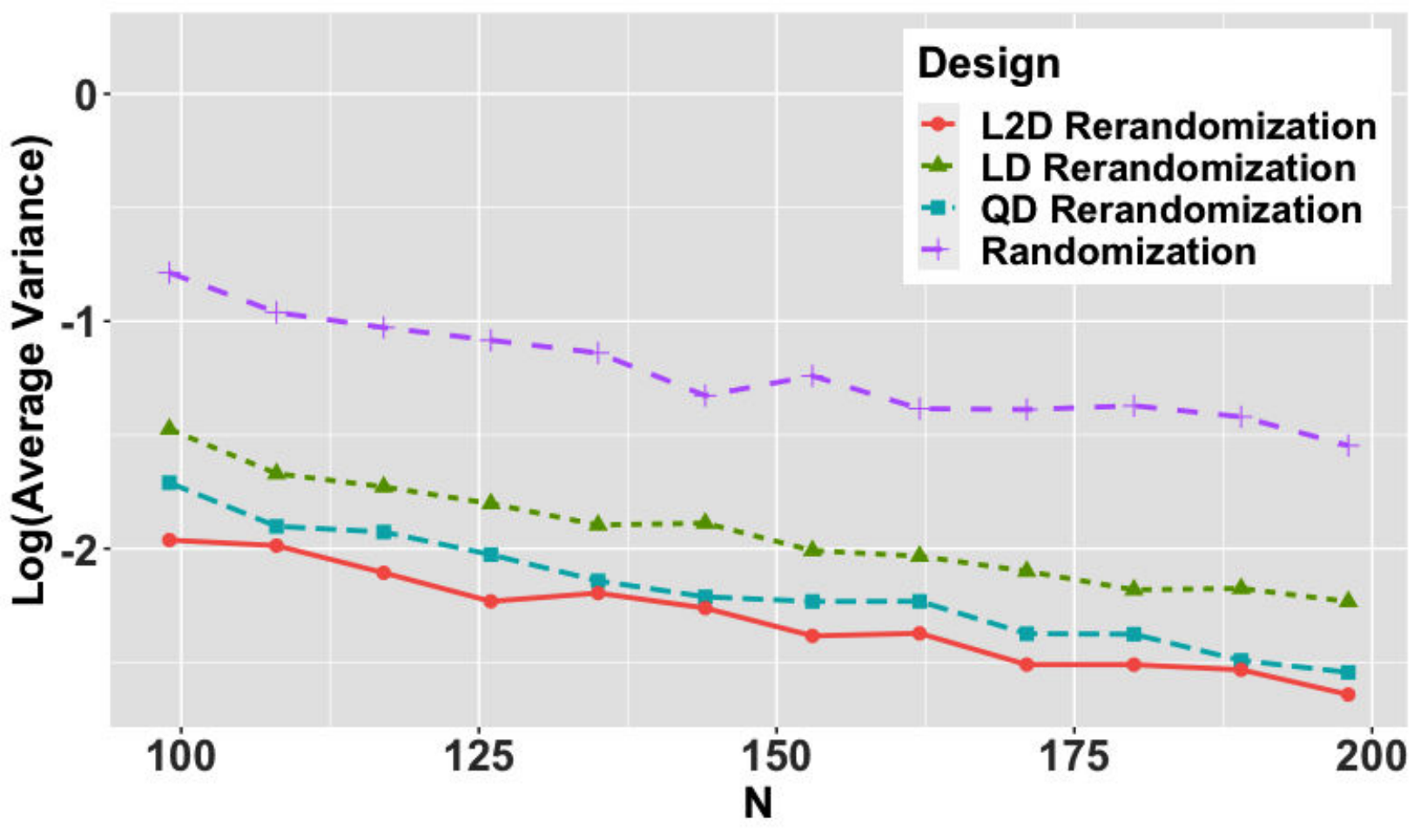}
\caption{$L=3$, Averaged Variance}
\end{subfigure}
\begin{subfigure}{.45\linewidth}
\includegraphics[width=\linewidth]{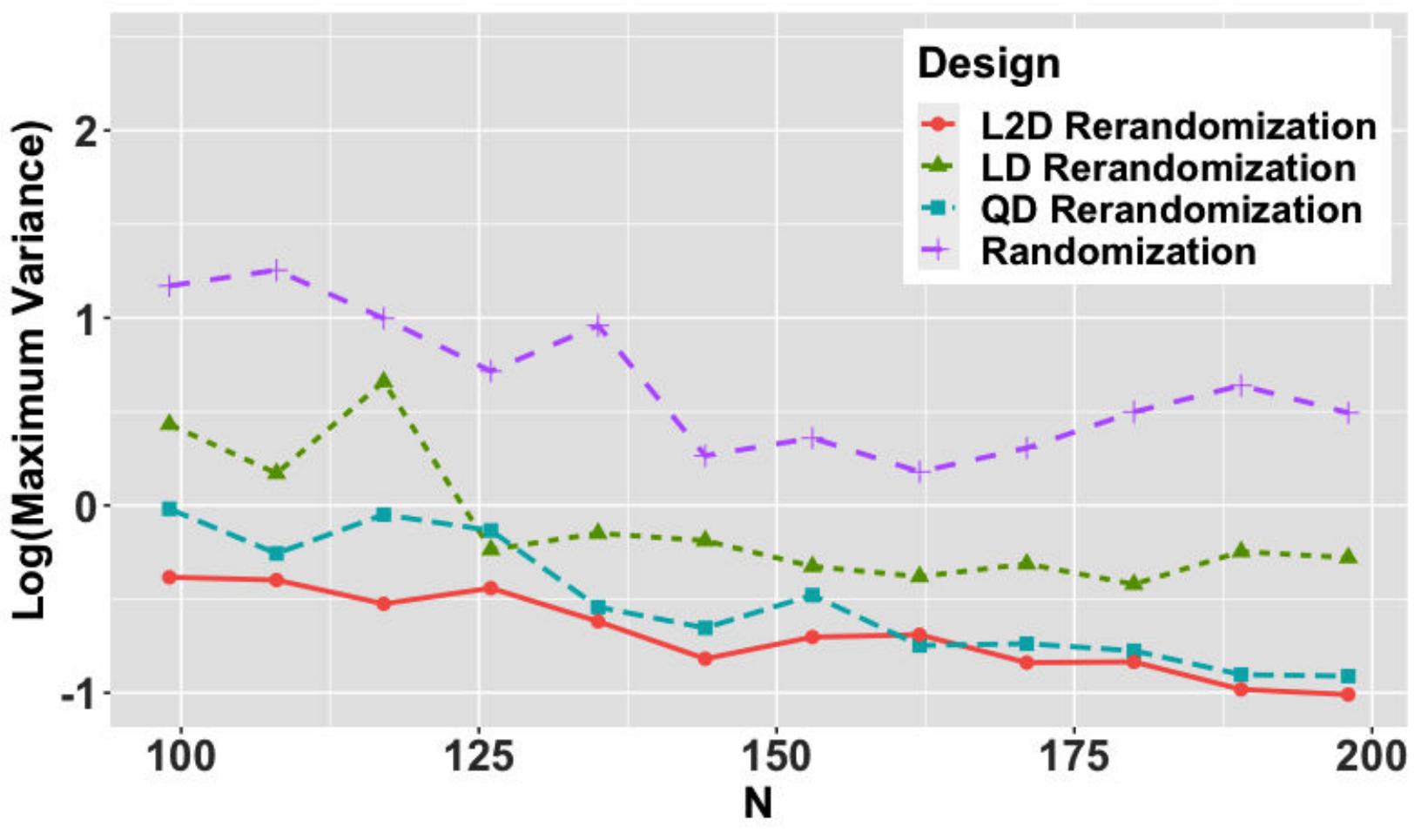}
\caption{$L=3$, Worst Variance}
\end{subfigure}
\caption{The average and worst-case estimated variance of the difference-in-means estimator using different random assignments for nonlinear $f(\bm z)$.}
\label{fig:2dnonpoly}
\end{figure}

For the rerandomization, thresholds $a_p$ are set to be $5\times 100\%$ percentile of the Gamma distribution for the linear discrepancy and the same percentile of the Chi-squared distribution for MD. 
For the $\mathcal{L}_2$ and quadratic kernel discrepancies, thresholds are estimated empirically using the $5\%$ sample percentile from 5000 random assignments. 
Figure \ref{fig:2dnonpoly} presents the comparison of the complete randomization and four rerandomizations in terms of the average and the worst-case estimated variance of the difference-in-means estimator.
Given the nonlinear nature of the relation function $f(\bm z)$, the rerandomization based on nonlinear kernel discrepancies - specifically, the $\mathcal{L}_2$ and quadratic discrepancy-based rerandomization outperform the other types. 
Among these two, the $\mathcal{L}_2$-based rerandomization yields slightly better results, because $f(\bm z)$ lies in the RKHS of the $\mathcal{L}_2$ kernel, but not in the RKHS induced by the quadratic kernel.

\section{Conclusion}\label{sec:end}

This paper has established kernel discrepancy as a powerful and unifying framework for rerandomization in controlled experiments. 
By moving beyond moment-matching criteria like the Mahalanobis distance, our approach ensures balance across the entire distribution of covariates, thereby offering a more robust and model-free rerandomization strategy.

We showed that the kernel discrepancy directly governs the upper bound of the variance for the difference-in-means estimator, providing a clear and principled rationale for its use as a balancing criterion.
A key enabling property of this framework is its inherent scalability to any number of treatment groups ($L\geq 2$), effectively overcoming the computational and theoretical limitations of pairwise multi-group extensions prevalent in the literature.

Building on this, we introduced a composite rerandomization strategy that makes balanced assignment feasible for large factorial experiments. 
By recursively applying two-group rerandomizations within a tree structure, this method achieves global covariate balance without the prohibitive computational cost of a direct $L$-group search.

Our numerical studies validated the practical utility of the proposed method. Rerandomization using kernel discrepancy consistently reduced estimator variance compared to complete randomization. Furthermore, the comparison of different kernels provided practical guidance: while the linear kernel is optimal for linear relationships, the $\mathcal{L}_2$-discrepancy offers a robust default under model uncertainty, effectively controlling the worst-case variance for a broader class of functions.

In summary, the kernel discrepancy framework provides a versatile and theoretically grounded enhancement to experimental design. 
It improves estimation precision by guaranteeing distributional balance across treatment groups, all while preserving the validity of randomization-based inference. Future research could extend this framework to dynamic experimental settings, integrate it with machine learning models for outcome adaptation, and develop scalable optimization techniques for very large-scale or sequential experiments.

\bibliography{Ref_ABtesting}

\newpage
\doublespacing
\begin{center}
{\Large\bf Supplementary Material}
\end{center}

\setcounter{figure}{0}
\setcounter{table}{0}
\setcounter{lemma}{0}
\setcounter{theorem}{0}
\setcounter{proposition}{0}

\makeatletter 
\renewcommand{\thefigure}{S\@arabic\c@figure}
\renewcommand{\thetable}{S\@arabic\c@table}
\renewcommand{\thelemma}{S\@arabic\c@lemma}
\renewcommand{\theproposition}{S\@arabic\c@proposition}
\renewcommand{\thetheorem}{S\@arabic\c@theorem}
\makeatother

\section*{S1. Proofs and Derivations}

\subsection*{Proof of Proposition \ref{prop:mean-var}}

\begin{proof}
The expectation is with respect to the response variable and the random assignment.
According to rerandomization procedure, $\E[\mathbbm{1}(X_i=q)|\mZ,\phi(\mZ,\vX)=1]=1/L$, and $X_i$ is conditionally independent of $Y_i$.
Thus,
\begin{align*}
&\E[\hat{\alpha}_{q}|\mZ,\phi(\mZ,\vX)=1]\\
=&\E\left[\left.\frac{\sum_{i=1}^N Y_{obs,i}\mathbbm{1}(X_i=q)}{\sum_{i=1}^N \mathbbm{1}(X_i=q)}\right|\mZ,\phi(\mZ,\vX)=1\right]
=\E\left[\left.\frac{1}{n}\sum_{i=1}^N Y_i(q)\mathbbm{1}(X_i=q)\right|\mZ,\phi(\mZ,\vX)=1\right]\\
=&\frac{1}{n}\sum_{i=1}^N\E[Y_i(q)|\mZ]\E[\mathbbm{1}(X_i=q)|\mZ,\phi(\mZ,\vX)=1]
=\frac{1}{n}\sum_{i=1}^N (\alpha_q+f(\vz_i))\frac{1}{L}= \alpha_q+\frac{1}{N}\sum_{i=1}^N f(\bm z_i),
\end{align*}
and it leads to the unbiasness of $\hat{\alpha}_q-\hat{\alpha}_{q'}$.
\cite{morgan2012rerandomization} had a similar proof for this result for $L=2$ case without any model assumption.

The variance of the difference-in-means estimator is
\begin{align*}
& \var[\hat{\alpha}_{q}-\hat{\alpha}_{q'}|\mZ,\phi(\mZ,\vX)=1]
=\E[(\hat{\alpha}_{q}-\hat{\alpha}_{q'}-(\alpha_q-\alpha_{q'}))^2|\mZ,\phi(\mZ,\vX)=1]\\
=& \E\left[\left.\left(\frac{1}{n}\sum_{i=1}^N Y_i(q)\mathbbm{1}(X_i=q)-\frac{1}{n}\sum_{i=1}^N Y_{i}(q')\mathbbm{1}(X_i=q')-(\alpha_q-\alpha_{q'})\right)^2\right|\mZ,\phi(\mZ,\vX)=1\right]\\
=&  \E\left[\left(\frac{1}{n}\sum_{i=1}^N f(\bm z_i)\mathbbm{1}(X_i=q)+\frac{1}{n}\sum_{i=1}^N\epsilon_i\mathbbm{1}(X_i=q)\right.\right.\\
&\left.\left.\left.-\frac{1}{n}\sum_{i=1}^N f(\bm z_i)\mathbbm{1}(X_i=q')-\frac{1}{n}\sum_{i=1}^N\epsilon_i\mathbbm{1}(X_i=q')\right)^2\right|\mZ,\phi(\mZ,\vX)=1\right]\\
=& \E\left[\left.\left(\int f(\vz)\dif F_{q}(\vz)-\int f(\vz)\dif F_{q'}(\vz)+\frac{1}{n}\sum_{i=1}^N\epsilon_i(\mathbbm{1}(X_i=q)-\mathbbm{1}(X_i=q'))\right)^2\right|\mZ,\phi(\mZ,\vX)=1\right]\\
=& \E\left[\left.\left(\int f(\vz)\dif F_{q}(\vz)-\int f(\vz)\dif F_{q'}(\vz)\right)^2\right|\mZ,\phi(\mZ,\vX)=1\right]+\frac{2\sigma^2}{n^2}\E(\sum_{i=1}^N\mathbbm{1}(X_i=q)|\mZ,\bm \phi(\mZ, \vX)=1)\\
=& \E\left[\left.\left(\int f(\vz)\dif F_{q}(\vz)-\int f(\vz)\dif F_{q'}(\vz)\right)^2\right|\mZ,\phi(\mZ,\vX)=1\right]+ \frac{2\sigma^2}{n}.
\end{align*}
\end{proof}

\subsection*{Proof of Theorem \ref{thm:mse}}
\begin{proof}
By inequality \eqref{eqn:integralbound}, for the $F_q$ and $F_{q'}$ resulted from an arbitrary randomization
\begin{equation}\label{eq:singleupper}
\left(\int f(\vz)\dif F_{q}(\vz)-\int f(\vz)\dif F_{q'}(\vz)\right)^2\leq D^2\left(F_q,F_{q'};K\right)[V(f)]^2.
\end{equation}
Thus,
\begin{eqnarray*}
&&\var[\hat{\alpha}_{q}-\hat{\alpha}_{q'}|\mZ,\phi(\mZ,\vX)=1]\\
&=& \E\left[\left.\left(\int f(\vz)\dif F_{q}(\vz)-\int f(\vz)\dif F_{q'}(\vz)\right)^2\right|\mZ,\phi(\mZ,\vX)=1\right]+ \frac{2\sigma^2}{n}\\
&\leq& \E\left[\left.D^2\left(F_q,F_{q'};K\right)\left[V\left(f\right)\right]^2\right|\mZ,\phi(\mZ,\vX)=1\right]+\frac{2\sigma^2}{n}\\
&=& \E\left[\left.D^2\left(F_q,F_{q'};K\right)\right|\mZ,\phi(\mZ,\vX)=1\right]\left[V\left(f\right)\right]^2+\frac{2\sigma^2}{n}
\end{eqnarray*}
\end{proof}

\subsection*{Proof of Theorem \ref{thm:discequality} and Corollary \ref{cor:discequality}}

Proof of Theorem \ref{thm:discequality}
\begin{proof}
By \eqref{eq:discDefdiscrete},
\begin{align*}
&\sum_{q,q'=1, q<q'}^L D^2\left(F_q,F_{q'};K\right)\\
&=\sum_{q,q'=1, q< q'}^L\left(\frac{1}{n^2}\sum_{i,k=1}^N K(\vz_i,\vz_k)\mathbbm{1}(X_i=q)\mathbbm{1}(X_k=q)-\frac{2}{n^2}\sum_{i,k=1, i< k}^N K(\vz_i,\vz_k)\mathbbm{1}(X_i=q)\mathbbm{1}(X_k=q')\right.\\
&\left.+\frac{1}{n^2}\sum_{i,k=1}^N K(\vz_i,\vz_k)\mathbbm{1}(X_i=q')\mathbbm{1}(X_k=q')\right)\\
&= \frac{1}{n^2}\sum_{q,q'=1, q<q'}^L \sum_{i,k=1}^N K(\vz_i,\vz_k)[\mathbbm{1}(X_i=q)\mathbbm{1}(X_k=q)-\mathbbm{1}(X_i=q)\mathbbm{1}(X_k=q')+\mathbbm{1}(X_i=q')\mathbbm{1}(X_k=q')]\\
&= \frac{1}{n^2}\sum_{i,k=1}^N K(\vz_i,\vz_k)\sum_{q,q'=1, q<q'}^L [\mathbbm{1}(X_i=q)\mathbbm{1}(X_k=q)-\mathbbm{1}(X_i=q)\mathbbm{1}(X_k=q')+\mathbbm{1}(X_i=q')\mathbbm{1}(X_k=q')].
\end{align*}
For any pair of $X_i, X_k$, it is easy to obtain
\begin{align*}
\sum_{q,q'=1, q<q'}^L &[\mathbbm{1}(X_i=q)\mathbbm{1}(X_k=q)-\mathbbm{1}(X_i=q)\mathbbm{1}(X_k=q')+\mathbbm{1}(X_i=q')\mathbbm{1}(X_k=q')] \\
& = \left\{\begin{array}{ll}
L-1,\,\,\,\,& \text{if }X_k=X_i\\
-1, & \text{if } X_k\neq X_i.
\end{array}\right.
\end{align*}
On the other hand, we can derive 
\begin{align*}
&\sum_{q=1}^L D^2(F_q,F;K)\\
= & \sum_{q=1}^L\left(\frac{1}{n^2}\sum_{i,k=1}^N K(\vz_i, \vz_k)\mathbbm{1}(X_i=q)\mathbbm{1}(X_k=q)-\frac{1}{nN}\sum_{i,k=1}^N K(\vz_i,\vz_k)\mathbbm{1}(X_i=q)+\frac{1}{N^2}\sum_{i,k=1}^N K(\vz_i,\vz_k)\right)\\
=& \frac{1}{n^2L}\sum_{i,k=1}^N K(\vz_i,\vz_k)\sum_{q=1}^L\left[L\mathbbm{1}(X_i=q)\mathbbm{1}(X_k=q)-2\mathbbm{1}(X_i=q)+1/L)\right],
\end{align*}
where
\begin{align*}
\sum_{q=1}^L\left[L\mathbbm{1}(X_i=q)\mathbbm{1}(X_k=q)-2\mathbbm{1}(X_i=q)+1/L)\right] = \left\{\begin{array}{ll}
L-1,\,\,\,\,& X_k=X_i\\
-1, & X_k\neq X_i.
\end{array}\right.
\end{align*}
Therefore, $\sum_{q=1}^L D^2(F_q,F;K) = \frac{1}{L}\sum_{q<q', q,q'\in\{1,\ldots,L\}} D^2\left(F_q,F_{q'};K\right)$.
\end{proof}

\noindent Proof of Corollary \ref{cor:discequality}
\begin{proof}
Following the similar proof of Theorem \ref{thm:discequality}, it can be shown that
$D^2(F_1,F;K) = D^2(F_2,F;K)$. 
Also based on Theorem \ref{thm:discequality},
$D^2(F_1,F;K)+D^2(F_2,F;K) = \frac{1}{2}D^2(F_1,F_2;K)$. Hence the result in the corollary. 
\end{proof}

\subsection*{Proof of Theorem \ref{thm:overallbound}}
\begin{proof}
By \eqref{ineqn:discrepancy},
\begin{eqnarray*}
&&\sum_{q<q',q,q'\in\{1,\ldots,L\}}\var[\hat{\alpha}_{q}-\hat{\alpha}_{q'}|\mZ,\phi(\mZ,\vX)=1]\\
&\leq& \sum_{q<q',q,q'\in\{1,\ldots,L\}}\left\{\E\left[\left.D^2\left(F_q,F_{q'};K\right)\right|\mZ,\phi(\mZ,\vX)=1\right]\left[V\left(f\right)\right]^2+\frac{2\sigma^2}{n}\right\}\\
&=& \E\left.\left[\sum_{q<q',q,q'\in\{1,\ldots,L\}}D^2\left(F_q,F_{q'};K\right)\right|\mZ,\phi(\mZ,\vX)=1\right]\left[V\left(f\right)\right]^2+\frac{\sigma^2L(L-1)}{n}\\
&=& \E\left.\left[\sum_{q=1}^L D^2(F_q,F;K)\right|\mZ,\phi(\mZ,\vX)=1\right]L\left[V\left(f\right)\right]^2+\frac{\sigma^2L(L-1)}{n}\\
&\leq & a_{p}L\left[V\left(f\right)\right]^2+\frac{\sigma^2L(L-1)}{n}
\end{eqnarray*}
\end{proof}

\subsection*{Proof of Proposition \ref{prop:discrelation}}
\begin{proof}
Let $\vt_i$'s denote the covariates of group $q$ and $\vx_i$'s the covariates of group $q'$. 
In this proof, we do not require the assignment to be balanced. 
Denote $n_1=\sum_{i=1}^N\sum \mathbbm{1}(X_i=q)$ and $n_2=\sum_{i=1}^N\sum \mathbbm{1}(X_i=q')$. 
In other words, $n_1$ and $n_2$ are the numbers of test units in group $q$ and $q'$, respectively. 
Following the definition of discrepancy, 
\begin{eqnarray*}
D^2(F_q,F_{q'};\linearK) &=& \frac{1}{n_1^2}  \sum_{i,j=1}^{n_1} \linearK(\vt_i,\vt_j) - \frac 2{n_1n_2} \sum_{i,j=1}^{n}  \linearK(\vt_i,\vx_j)  + \frac{1}{n_2^2}  \sum_{i,j=1}^{n_2} \linearK(\vx_i,\vx_j)\\
&=& \frac{1}{n_1^2}  \sum_{i,j=1}^{n_1}\sum_{k=1}^d t_{ik}t_{jk}  - \frac 2{n_1n_2} \sum_{i=1}^{n_1}\sum_{j=1}^{n_2}\sum_{k=1}^d x_{ik}t_{jk}  + \frac{1}{n_2^2}  \sum_{i,j=1}^{n_2}\sum_{k=1}^d x_{ik}x_{jk}\\
&=& \sum_{k=1}^d \left[\frac{1}{n_1^2}  \sum_{i,j=1}^{n_1}t_{ik}t_{jk}  - \frac 2{n_1n_2} \sum_{i=1}^{n_1}\sum_{j=1}^{n_2} t_{ik}x_{jk}  + \frac{1}{n_2^2}  \sum_{i,j=1}^{n_2} x_{ik}x_{jk}\right]\\
&=& \sum_{k=1}^d \left[\frac{(\sum_{i=1}^{n_1} t_{ik})^2}{n_1^2}   - \frac {2(\sum_{i=1}^{n_1} t_{ik})(\sum_{j=1}^{n_2} x_{jk})}{n_1n_2}  + \frac{(\sum_{i=1}^{n_2} x_{ik})^2}{n_2^2}  \right]\\
&=& \sum_{k = 1}^d\left(\frac{\sum_{i=1}^{n_1} t_{ik}}{n_1}-\frac{\sum_{i=1}^{n_2} x_{ik}}{n_2}\right)^2=\sum_{k = 1}^d\left(\bar{Z}_k^{(q)}-\bar{Z}_k^{(q')}\right)^2.
\end{eqnarray*}
The last equation is due to the newly introduced notation at the beginning of Section \ref{subsec:linear}. 
The same proof can be used to obtain $D^2(F_q, F; \linearK)$. 
\end{proof}

\subsection*{Lemma \ref{lem:correlation} and Proof}

\begin{lemma}\label{lem:correlation}
Denote the columns of $\mZ$ as $\bm Z_1,...,\bm Z_d$.
Assume that the finite population variances of covariates have been standardized, i.e., $\nu_j=\frac{1}{N-1}\sum_{i=1}^N(z_{ij}-\bar{Z}_j)^2 = 1$, $j=1,\cdots,d$, where $\bar{Z}_j$ is the mean of $\bm Z_j$. Denote $\bar{Z}_k^{(q)} = \frac{\sum_{i=1}^N z_{ik}\mathbbm{1}(X_i=q)}{n}, q=1,...L, k=1,...,d$ as the mean of the $k$th covariate in group $q$. Then, under the completely randomization with equal group size $n_1=\cdots=n_L=n=\frac{N}{L}$, for $k,k'\in\{1,...,d\}, q,q'\in\{1,...,L\}$,
\begin{equation}\label{eqn:correlation}
\corr\left(\bar{Z}_k^{(q)},\bar{Z}_{k'}^{(q')}\right)= \left\{\begin{array}{ll}
\corr(\bm Z_k,\bm Z_{k'}) ,\,\,\,& k\neq k', q=q'\\
-\frac{1}{L-1}\corr(\bm Z_k,\bm Z_{k'}), &  k\neq k',q\neq q'.\\
\end{array}\right.
\end{equation}

\begin{proof}
Under completely randomization with equal group size, it is obvious that $\mathbbm{1}(X_i=q) \sim Bernoulli(1/L)$ and $\sum_{i=1}^N \mathbbm{1}(X_i=q) = n$, for $i=1,...,N$, $q=1,...,L$. Thus,
$$\E\left[\mathbbm{1}(X_i=q)\mathbbm{1}(X_{i'}=q')\right] = \left\{\begin{array}{ll}
\frac{1}{L},\,\,\,& i = i',q = q\\
0, & i=i',q\neq q'\\
\frac{\binom{N-2}{n-2}}{\binom{N}{n}} = \frac{N-L}{L^2(N-1)}, & i\neq i', q=q'\\
\frac{\binom{N-2}{n-1}\binom{N-n-1}{n-1}}{\binom{N}{n}\binom{N-n}{n}} = \frac{N}{L^2(N-1)}, & i\neq i', q\neq q'
\end{array}\right..$$
Thus, for $k\neq k'$,
\begin{eqnarray*}
\E\left(\bar{Z}_k^{(q)}\bar{Z}_{k'}^{(q)}\right) &=& \frac{1}{n^2} \E\left[\left(\sum_{i=1}^Nz_{ik}\mathbbm{1}(X_i=q)\right)\left(\sum_{j=1}^Nz_{jk'}\mathbbm{1}(X_i=q)\right)\right]\\
&=& \frac{1}{n^2}\E\left[\sum_{i=1}^N z_{ik}z_{ik'}\mathbbm{1}(X_i=q)\mathbbm{1}(X_i=q)+\mathop{\sum^{N}\sum^{N}}_{i=1\  i\neq j\ j=1} z_{ik}z_{jk'}\mathbbm{1}(X_i=q)\mathbbm{1}(X_j=q)\right]\\
&=& \frac{1}{n^2} \left(\frac{1}{L}\sum_{i=1}^N z_{ik}z_{ik'}+\frac{N-L}{L^2(N-1)}\mathop{\sum^{N}\sum^{N}}_{i=1\  i\neq j\ j=1} z_{ik}z_{jk'}\right),
\end{eqnarray*}
and since $\E(\bar{Z}_k^{(q)}) = \bar{Z}_k$,
\begin{eqnarray*}
\cov\left(\bar{Z}_k^{(q)},\bar{Z}_{k'}^{(q)}\right) &=& \E\left(\bar{Z}_k^{(q)}\bar{Z}_{k'}^{(q)}\right) - \E\left(\bar{Z}_k^{(q)}\right)\E\left(\bar{Z}_{k'}^{(q)}\right)\\
&=& \frac{1}{n^2} \left(\frac{1}{L}\sum_{i=1}^N z_{ik}z_{ik'}+\frac{N-L}{L^2(N-1)}\mathop{\sum^{N}\sum^{N}}_{i=1\  i\neq j\ j=1} z_{ik}z_{jk'}\right) - \frac{\sum_{j=1}^Nz_{jk}}{N}\frac{\sum_{j=1}^Nz_{jk'}}{N}\\
&=& \frac{1}{N^2}\left(L\sum_{i=1}^N z_{ik}z_{ik'}+\frac{N-L}{(N-1)}\mathop{\sum^{N}\sum^{N}}_{i=1\  i\neq j\ j=1} z_{ik}z_{jk'}-\sum_{i=1}^N z_{ik}z_{ik'}-\mathop{\sum^{N}\sum^{N}}_{i=1\  i\neq j\ j=1} z_{ik}z_{jk'}\right)\\
&=& \frac{L-1}{N^2}\left(\sum_{i=1}^N z_{ik}z_{ik'}-\frac{1}{(N-1)}\mathop{\sum^{N}\sum^{N}}_{i=1\  i\neq j\ j=1} z_{ik}z_{jk'}\right).
\end{eqnarray*}
By the sampling theory, $\var\left(\bar{Z}_k^{(q)}\right) = \left(\frac{1}{N/L}-\frac{1}{N}\right) = \frac{L-1}{N}$. As a result,
$$
\corr\left(\bar{Z}_k^{(q)},\bar{Z}_{k'}^{(q)}\right) = \frac{\cov\left(\bar{Z}_k^{(q)},\bar{Z}_{k'}^{(q)}\right)}{\sqrt{\var\left(\bar{Z}_k^{(q)}\right)\var\left(\bar{Z}_{k'}^{(q)}\right)}}
= \frac{1}{N}\left(\sum_{i=1}^N z_{ik}z_{ik'}-\frac{1}{N-1}\mathop{\sum^{N}\sum^{N}}_{i=1\  i\neq j\ j=1}z_{ik}z_{jk'}\right).$$
Finally, since the finite population variance $\nu_j=\frac{1}{N-1}\sum_{i=1}^N(z_{ij}-\bar{Z}_j)^2 = 1$ is standardized,
\begin{eqnarray*}
\corr(\bm Z_k,\bm Z_{k'}) &=&
\frac{\sum_{i=1}^N z_{ik}z_{ik'}-N\frac{\sum_{i=1}^Nz_{ik}}{N}\frac{\sum_{i=1}^Nz_{ik'}}{N}}{N-1}\\
& =&  \frac{1}{N}\left(\sum_{i=1}^N z_{ik}z_{ik'}-\frac{1}{N-1}\mathop{\sum^{N}\sum^{N}}_{i=1\  i\neq j\ j=1}z_{ik}z_{jk'}\right) =\corr\left(\bar{Z}_k^{(q)},\bar{Z}_{k'}^{(q)}\right).
\end{eqnarray*}
For $k\neq k'$ and $q\neq q'$, following similar calculation,
\begin{eqnarray*}
\E\left(\bar{Z}_k^{(q)}\bar{Z}_{k'}^{(q')}\right) &=& \frac{1}{n^2} \E\left[\left(\sum_{i=1}^Nz_{ik}\mathbbm{1}(X_i=q)\right)\left(\sum_{j=1}^Nz_{jk'}\mathbbm{1}(X_i=q')\right)\right]\\
&=& \frac{1}{n^2} \left(\frac{N}{L^2(N-1)}\mathop{\sum^{N}\sum^{N}}_{i=1\  i\neq j\ j=1} z_{ik}z_{jk'}\right).
\end{eqnarray*}
The rest of the derivation would be almost the same as above, and we have
$$\corr\left(\bar{Z}_k^{(q)},\bar{Z}_{k'}^{(q')}\right) = -\frac{1}{L-1}\corr(\bm Z_k,\bm Z_{k'}),$$
for $k\neq k',q\neq q'$.
\end{proof}
\end{lemma}

\subsection*{Proof Theorem \ref{thm:discdistrfinite}}

\begin{proof} By the Central Limit Theorem of finite population \citep{hajek1960limiting}, for $k=1,...,d$, $$\sqrt{N}\left(\bar{Z}_k^{(1)}-\bar{Z}_k\right)\stackrel{\text{d}}{\rightarrow} N(0,1), $$
and $N\left(\bar{Z}_k^{(1)}-\bar{Z}_k\right)^2\stackrel{\text{d}}{\rightarrow} \chi^2_1$,  as $N\rightarrow\infty$. 
Here $\stackrel{\text{d}}{\rightarrow}$ stands for convergence in distribution.

Denote the vector $\bm W = [W_1,...,W_d]$, where $W_k = \sqrt{N}\left(\bar{Z}_k^{(1)}-\bar{Z}_k\right)$, $k=1,...,d$. 
Then, by Lemma \ref{lem:correlation}, $\corr(W_j,W_k)=\corr\left(\bar{Z}_j^{(1)},\bar{Z}_k^{(1)}\right)=\rho_{jk}$, $j\neq k, j,k=1,...,d$.  As a result, $(W_j,W_k) \stackrel{\text{d}}{\rightarrow} N\left(\mathbf{0},\left[\begin{array}{ll}1\,\,\,\,& \rho_{jk}\\\rho_{jk} & 1\end{array}\right]\right)$.
Then, $\lim\limits_{N\rightarrow \infty}\cov(W_j^2,W_k^2) = 2\rho^2_{jk}$ and $\lim\limits_{N\rightarrow\infty}\corr(W_j^2,W_k^2) =  \rho^2_{jk}$. Based on the distribution of the sum of correlated $\chi^2$ random variables \citep{paris2011note}, the asymptotic distribution of $N\times D^2\left(F_1,F;\linearK\right)= N \sum_{k=1}^d\left(\bar{Z}_k^{(1)}-\bar{Z}_k\right)^2 = \sum_{k=1}^dW_k^2$ has the CDF function
$$F_{ND^2}(y) = \frac{y^{d/2}}{\det(\mA^{1/2})\Gamma(1+d/2)}\times \Phi_2^{(d)}\left(\frac{1}{2},...,\frac{1}{2};1+\frac{d}{2};-\frac{y}{\lambda_1},...,-\frac{y}{\lambda_d}\right),\,\,\,\,y>0,$$
where $\Phi_2^{(d)}$, $\{\lambda_i\}_{i=1}^d$, and $\mA$ are defined in the theorem. 
\end{proof}

\subsection*{Proof of Theorem \ref{thm:discfiniteapprox}}

\begin{proof} Based on the approximate distribution of the sum of correlated Chi-square random variables \citep{ferrari2019note}, $N \sum_{k=1}^d\left(\bar{Z}_k^{(1)}-\bar{Z}_k\right)^2  = N D^2\left(F_1,F;\linearK\right)\stackrel{\text{d}}{\rightarrow} \Gamma\left(\frac{d}{u},u\right)$, where $u = 2\left(1+\frac{2}{d}\sum\limits_{j\neq k}^d\rho^2_{jk}\right)$ and $\rho_{ij}$ is the correlation between $\bm Z_j$ and $\bm Z_k$. 
By Corollary \ref{cor:discequality}, the result on $N\times D^2(F_1,F_2;\linearK)$ follows.
\end{proof}

\subsection*{Proof of Theorem \ref{thm:linearL3approx}}

\begin{proof}
Define an $(L-1)\times d$ matrix $\mW$, whose $(q,k)$-th entry is 
\[W_{qk} = \frac{\bar{Z}_k^{(q)}-\bar{Z}_k}{\sqrt{\frac{L-1}{N}}}, \text{ for } q=1,...,L-1, k=1,...,d.\] 
For the $L$-th treatment group, the mean of the $k$th covariate is 
\[
\bar{Z}_k^{(L)} = \frac{L}{N}\left(N\bar{Z}_k-\frac{N}{L}\sum_{q=1}^{L-1}\bar{Z}_{k}^{(q)}\right),\]
and 
\[
\bar{Z}_k^{(L)}-\bar{Z}_k = (L-1)\bar{Z}_k-\sum_{q=1}^{L-1}\bar{Z}_k^{(q)} = \sum_{q=1}^{L-1} \left(\bar{Z}_k-\bar{Z}_k^{(q)}\right) = -\sum_{q=1}^{L-1}\sqrt{\frac{L-1}{N}}W_{qk}.
\]
Thus, the total squared discrepancy can be rewritten as
\begin{eqnarray*}
\sum_{q=1}^L D^2(F_q,F;\linearK) &=& \sum_{q=1}^L \sum_{k=1}^d\left(\bar{Z}_k^{(q)}-\bar{Z}_k\right)^2 \\
&=& \sum_{k=1}^d\left[\sum_{q=1}^{L-1}\frac{L-1}{N}W_{qk}^2+ \left(\sum_{q=1}^{L-1}\sqrt{\frac{L-1}{N}}W_{qk}\right)^2\right]\\
&=& \sum_{k=1}^d\frac{L-1}{N}\left(2\sum_{q=1}^{L-1}W^2_{qk}+\mathop{\sum^{L-1}\sum^{L-1}}_{q=1\  q\neq q'\ q'=1} W_{qk}W_{q'k}\right).
\end{eqnarray*}
Denote 
\[
D_k = \frac{L-1}{N}\left(2\sum_{q=1}^{L-1}W^2_{qk}+\mathop{\sum^{L-1}\sum^{L-1}}\limits_{q,q'=1\  q\neq q'} W_{qk}W_{q'k}\right)
\]
and the columns of $\mW$ as $\bm W_k$, $k=1,...,d$. 
Then, $D_k$ can be written as
$D_k = \frac{L-1}{N}\bm W_k^\top\mA\bm W_k,$
where 
\[
\mA = \begin{pmatrix}
2 & 1 & \cdots & 1\\
1 & 2 & \cdots &1\\
\vdots & \vdots & \ddots & \vdots\\
1 & 1 & \cdots &2
\end{pmatrix}_{(L-1)\times (L-1)}.
\]

As shown in \citep{li2017general}, $\bm W_k\rightarrow N(0,\Sigma),\,\,\,\,k=1,...,d$, as $N\rightarrow \infty$, where 
\[
\Sigma = \begin{pmatrix}
1 & -\frac{1}{L-1} & \cdots & -\frac{1}{L-1}\\
-\frac{1}{L-1} & 1 &\cdots & -\frac{1}{L-1}\\
\vdots & \vdots & \ddots & \vdots\\
-\frac{1}{L-1} & -\frac{1}{L-1} &\cdots &1
\end{pmatrix}_{(L-1)\times (L-1)}.
\]
We can standardize $\bm W_k$ as ${\bm SW}_k = \Sigma^{-1/2}\bm W_k\stackrel{\text{d}}{\rightarrow} N(0, \mI)$, where $\mI$ is the identity matrix.

We need the following lemma to proceed with the proof of Theorem \ref{thm:linearL3approx}.
\begin{lemma}\label{lem:sigma}
$\Sigma^{1/2}\mA\Sigma^{1/2} = \frac{L}{L-1}\mI,$
where $\mI$ is the identity matrix.
\end{lemma}
\begin{proof} Denote the $(L-1)\times (L-1)$ unit matrix with all ones as 
$$\mU = \begin{pmatrix}
1 & 1 & \cdots &1\\
\vdots  & \vdots & \ddots & \vdots\\
1 & 1 &\cdots& 1
\end{pmatrix}_{(L-1)\times (L-1)}.$$ 
Then $\mA = \mI+\mU$.
So, $\Sigma^{1/2}\mA\Sigma^{1/2} = \Sigma+\Sigma^{1/2}\mU\Sigma^{1/2}$.

Since $\frac{\mU}{L-1}$ is idempotent and $\mU$ and $\Sigma^{1/2}$ are positive-semi definite, 
\begin{align*}
\Sigma^{1/2}\mU\Sigma^{1/2}& = (L-1)\Sigma^{1/2}\frac{\mU}{L-1}\frac{\mU}{L-1}\Sigma^{1/2} \\
& = (L-1)\frac{\mU}{L-1}\Sigma\frac{\mU}{L-1} = \begin{pmatrix}
\frac{1}{L-1} & \frac{1}{L-1} & \cdots & \frac{1}{L-1}\\
\vdots& \vdots &\ddots & \vdots\\
\frac{1}{L-1} & \frac{1}{L-1} &\cdots & \frac{1}{L-1}
\end{pmatrix}_{(L-1)\times (L-1)},
\end{align*}
a matrix whose entries are $\frac{1}{L-1}$. 
Thus, $\Sigma^{1/2}\mA\Sigma^{1/2} = \frac{L}{L-1}\mI$.
\end{proof}

Based on the Lemma \ref{lem:sigma},
$D_k = \frac{L-1}{N}\bm W_k^\top\mA\bm W_k = \frac{L-1}{N}{\bm SW}_k^{\top}\Sigma^{1/2}\mA\Sigma^{1/2}{\bm SW}_k =   \frac{L}{N}\sum_{q=1}^{L-1}{\bm SW}_{qk}^2$. $SW_{qk}$ are the elements in the following matrix
$$\bm SW = \begin{bmatrix}
SW_{11} & SW_{12} &\cdots & SW_{1d}\\
\vdots & \vdots & \vdots & \vdots\\
SW_{(L-1)1} & SW_{(L-1) 2} &\cdots & SW_{(L-1) d}
\end{bmatrix},$$
where each column are independent standard normal random variables asymptotically.

The following Lemma \ref{lem:SW} shows the correlation properties between the elements in matrix $SW$.
\begin{lemma}\label{lem:SW}
$$\corr(SW_{qk},SW_{q'k'}) = \left\{\begin{array}{ll}
1,\,\,\,& q=q', k=k'\\
\corr(\bm Z_k, \bm Z_{k'}), &q=q',k\neq k'\\
0 & \text{otherwise}
\end{array}\right.$$
That is, $\corr(SW_{qk},SW_{qk'})=\corr(\bm Z_k, \bm Z_{k'})=\rho_{kk'}$ for $k\neq k'$, and the correlation is zero between the elements in the same column or elements with both different column and row indexes.
\end{lemma}
\begin{proof}
We first derive the explicit formula for $SW_{qk}$.
The eigenvalues of $\Sigma$ are $\lambda_1 = \frac{1}{L-1}$ with multiplicity 1 and $\lambda_2 = \frac{L}{L-1}$ with multiplicity $L-2$. The eigenvector corresponding to $\lambda_1$ is $\bm v_1 = [1,1,\ldots,1]^\top$, and the eigenvectors for $\lambda_2$ are $\bm v_i =  e_1-e_i$, $i=2,\ldots,L-1$, where $e_i$ is the column vector of zeros except the $i$th element as 1.  Define $\mV = [\bm v_1 | \ldots | \bm v_{L-1}]$. Then,  $\mV^{-1}_{ii} = -\frac{L-2}{L-1}$, for $i=2,\ldots,  L-1$, $\mV^{-1}_{ij} = \frac{1}{L-1}$, for $i=j=1$ or $i\neq j, i,j=1,\ldots, L-1$. With eigenvalue decomposition, $(\Sigma^{-1/2})_{ii} = \frac{(\sqrt{L}-1)(\sqrt{L}+2)}{\sqrt{L(L-1)}}$, for $i=1,\ldots,L-1$, and $(\Sigma^{-1/2})_{ij} = \frac{\sqrt{L}-1}{\sqrt{L(L-1)}}$,  for $i,j=1,\ldots, L-1$ and $i\neq j$. Since
${\bm SW}_k = \Sigma^{-1/2}\bm W_k$, we have $SW_{qk} = \frac{\sqrt{L}-1}{\sqrt{L(L-1)}}\sum_{j=1}^{L-1} W_{jk}+\sqrt{\frac{L-1}{L}}W_{qk} = \frac{\sqrt{L}-1}{\sqrt{L-1}}\left(\frac{1}{\sqrt{L}}\sum_{j=1}^{L-1} W_{jk}+W_{qk}\right)$.

Since $W_{qk}$ is the standardized group mean, by Lemma \ref{lem:correlation}, it is easy to show that
\begin{eqnarray*}
\E(W_{qk},W_{q'k'}) & = &\cov(W_{qk},W_{q'k'}) =  \corr(W_{qk},W_{q'k'}) = \corr\left(\bar{Z}_k^{(q)},\bar{Z}_{k'}^{(q')}\right)\\
&=& \left\{\begin{array}{ll}
1, \,\,\,& q=q',k=k'\\
\corr(\bm Z_k,\bm Z_{k'}), & q=q',k\neq k'\\
-\frac{1}{L-1}, & q\neq q',k=k'\\
-\frac{1}{L-1}\corr(\bm Z_k,\bm Z_{k'}), & q\neq q',k\neq k'
\end{array}\right..
\end{eqnarray*}
Thus, when $q=q'$ and $k\neq k'$,
\begin{eqnarray*}
& &\corr(SW_{qk},SW_{q'k'}) = \E(SW_{qk}SW_{q'k'})\\
&=& \left(\frac{\sqrt{L}-1}{\sqrt{L-1}}\right)^2\E\left[\left(\frac{1}{\sqrt{L}}\sum_{j=1}^{L-1} W_{jk}+W_{qk}\right)\left(\frac{1}{\sqrt{L}}\sum_{j=1}^{L-1} W_{jk'}+W_{q'k'}\right)\right]= 0,
\end{eqnarray*}
by simple algebra.
The other correlations can be derived similarly.
\end{proof}

Now, we can rewrite the rerandomization criterion as
$$\sum_{q=1}^L D^2(F_q,F;\linearK) = \sum_{k=1}^d \frac{L}{N}\sum_{q=1}^{L-1}{\bm SW}_{qk}^2 =\frac{L}{N} \sum_{q=1}^{L-1}\sum_{k=1}^d {\bm SW}_{qk}^2.$$
By \citep{ferrari2019note}, the row sums, as the sum of correlated chi-square random variables,
$$\sum_{k=1}^d SW_{qk}^2\stackrel{\text{d}}{\rightarrow} \Gamma\left(\frac{d}{u},u\right),$$ where $u = 2\left(1+\frac{2}{d}\sum\limits_{j\neq k}^d\rho^2_{jk}\right)$ and $\rho_{jk}$ is the correlation between $\bm Z_j$ and $\bm Z_{k}$, for
$q=1,\ldots,L-1$. Then, $\sum_{q=1}^{L-1}\sum_{j=k}^d SW_{qk}^2$ is the sum of independent gamma random variables, so $ \sum_{q=1}^{L-1}\sum_{k=1}^d SW_{qk}^2\rightarrow \Gamma\left(\frac{(L-1)d}{u},u\right)$. By the scale property of gamma distribution
$$\sum_{q=1}^L D^2(F_q,F;\linearK)  \stackrel{\text{d}}{\rightarrow} \Gamma\left(\frac{(L-1)d}{u},\frac{Lu}{N}\right),$$
where $\frac{(L-1)d}{u}$ is the shape parameter, $\frac{Lu}{N}$ is the scale parameter, $u = 2\left(1+\frac{2}{d}\sum\limits_{j\neq k}^d\rho^2_{jk}\right)$, and $\rho_{jk}$ is the correlation between $\bm Z_j$ and $\bm Z_{k}$.
\end{proof}

\end{document}